# Cation vacancies mediate thermochemical water splitting with iron aluminates

Nathan J. Szymanski[1], Kent J. Warren[2], Alan W. Weimer[3], and Christopher J. Bartel[1,*]

## Abstract

Solar thermochemical water splitting enables hydrogen production by cycling metal oxides between reduced and oxidized states, typically through an oxygen vacancy mechanism. However, recent experimental work suggests that cation vacancies have a greater influence on the redox behavior of iron aluminate spinels used in water splitting. This remains debated, as calculations predict that such cation vacancies are thermodynamically unfavorable. In the current work, we show that Fe vacancies in $(Fe_{\zeta}Al_{1-\zeta})_3O_4$ become accessible only when facilitated by inversion between Fe and Al. This antisite disorder lowers the formation energy of octahedral Fe vacancies in Al-rich spinels ($\zeta = 1/3$) from over 3 eV to just 0.62 eV when one third of the cation sites are inverted, allowing high Fe vacancy concentrations under oxidizing conditions. This mechanism supports high $H_2$ yields up to 361 μmol/g, consistent with experimental observations. Our findings support the notion that solar thermochemical water splitting can occur through a cation vacancy mechanism. They also clarify how site inversion, vacancy energetics, and defect interactions each contribute to redox performance, offering general design principles for identifying and optimizing materials that operate through cation vacancy cycling.

[1] University of Minnesota, Department of Chemical Engineering and Materials Science, Minneapolis, MN 55455

[2] Solar Energy Engineering Laboratory, Department of Mechanical and Process Engineering, ETH Zürich, 8092 Zürich, Switzerland

[3] Department of Chemical and Biological Engineering, University of Colorado, Boulder, CO 80309

[*] Correspondence to cbartel@umn.edu

## Introduction

The conversion of solar energy into fuel *via* solar thermochemical water splitting (STWS) offers a promising route toward sustainable, carbon-free hydrogen generation.[1–4] In this process, a metal oxide is cyclically reduced and oxidized at high temperatures,[5] using concentrated solar radiation to separately drive oxygen release and water dissociation. Early work focused on stoichiometric redox pairs,[6] such as ZnO/Zn[7,8] and $Fe_3O_4$/FeO,[9,10] which undergo phase transformations during cycling and yield relatively high $H_2$ production. However, these systems often suffer from slow reaction kinetics and degradation caused by sintering.[11] To address these challenges, researchers have turned to nonstoichiometric oxides[12–16] – most notably, $CeO_{2-\delta}$ – which mediate redox reactions through reversible oxygen vacancy formation while maintaining structural integrity. Nevertheless, oxides like $CeO_{2-\delta}$ face limitations due to the extremely high temperatures needed to generate a sufficient concentration of oxygen vacancies ($\delta$) to enable high $H_2$ yields and effective oxidant (water-to-hydrogen) conversion.[17] These challenges have motivated research on alternative materials for STWS,[18] such as perovskites[19–24] and spinels,[25,26] which may offer greater compositional flexibility and higher vacancy concentrations at lower cycling temperatures.

Iron aluminate spinels, with the general formula $(Fe_{\zeta}Al_{1-\zeta})_3O_4$, have emerged as promising candidates for STWS owing to their structural stability, earth abundance, and flexible redox chemistry.[27–31] These materials adopt the same crystal structure as magnetite ($Fe_3O_4$), comprising a close-packed oxygen lattice with cations distributed between the tetrahedral and octahedral sites. Recent work suggests that, unlike traditional STWS materials where oxygen vacancies are the dominant defect species, redox in $(Fe_{\zeta}Al_{1-\zeta})_3O_4$ occurs predominantly through cation vacancy formation and annihilation.[30,32] Experimental support for this mechanism includes diffraction measurements by Al-Shankiti *et al.*, who observed counterintuitive lattice contraction in $FeAl_2O_4$ ($\zeta = 1/3$) during $Fe^{3+} \rightarrow Fe^{2+}$ reduction and attributed it to the presence of cation vacancies.[30] Warren *et al.* further showed through high-temperature thermochemical measurements that iron aluminates with high Al content ($\zeta < 1/2$) exhibit substantial cation deficiency under oxidizing conditions.[32] This reported behavior aligns with previous findings on $Fe_3O_4$, where Fe vacancies are the dominant defect species under oxidizing conditions, facilitating rapid cation diffusion within the material.[33]

Recent first-principles calculations by Millican *et al.* provide an important foundation for understanding the defect thermochemistry in iron aluminates.[34] Their study benchmarked several functionals and charge states, finding that oxygen vacancies paired with Fe/Al antisite defects are the most favorable complexes in $FeAl_2O_{4-\delta}$ under STWS-relevant conditions. They showed that this antisite-vacancy pairing increases the oxygen deficiency ($\delta$) by 3-4 orders of magnitude relative to isolated vacancies and also highlight the effect of cation off-stoichiometry ($\zeta$) on oxygen vacancies. However, the predicted oxygen deficiency ($\delta = 2 \times 10^{-3}$ under Fe-rich conditions at 1400 °C and $p_{O_2} = 10^{-4}$ atm) remains well below what is needed to fully account for experimental $H_2$ yields, suggesting that additional mechanisms may contribute during cycling.

In this work, we predict that cation vacancies play a major role in the STWS mechanism of iron aluminate spinels, but only when facilitated by cation inversion – that is, antisite disorder between the octahedral and tetrahedral sublattices. To investigate this mechanism, we studied three $(Fe_{\zeta}Al_{1-\zeta})_3O_4$ compositions with $\zeta$ = 1/3 ($FeAl_2O_4$), 1/2 ($Fe_{1.5}Al_{1.5}O_4$), and 1 ($Fe_3O_4$). Our DFT calculations were first benchmarked against known experimental trends, including site inversion in $FeAl_2O_4$ and defect concentrations in $Fe_3O_4$. We then examined how inversion affects the formation of cation vacancies in $(Fe_{\zeta}Al_{1-\zeta})_3O_4$, showing that antisite disorder dramatically lowers the energy cost of removing Fe from the octahedral sublattice. This allows the iron aluminates to achieve high Fe vacancy concentrations under typical STWS conditions. By calculating the equilibrium Fe vacancy concentration across varying $O_2$ partial pressure, we link these changes to $H_2$ yield and show reasonable agreement between calculation and experiment. Based on these findings, we provide guidance for the design of new STWS materials that operate *via* cation vacancies.

## Methods

### $Fe_3O_4$ vacancy calculations

To validate our computational approach, we first benchmarked it against the known defect chemistry in magnetite ($Fe_3O_4$). This well-studied material crystallizes in the cubic spinel structure above the Verwey transition temperature of 122 K.[35] Below this temperature, $Fe_3O_4$ adopts a monoclinic structure; however, our calculations assume cubic symmetry throughout, as this is the relevant phase under STWS conditions ($T > 1000$ °C). Furthermore, our DFT calculations indicate that Al substitution on the octahedral sublattice progressively stabilizes the cubic spinel over the monoclinic phase (Supplementary Figure 1), supporting the use of cubic models for the iron aluminate compositions studied in later sections.

This cubic spinel structure, with generic composition $AB_2O_4$, features a close-packed oxygen lattice with one $A$ ion ($Fe^{2+}$) per formula unit on the tetrahedral sites and two $B$ ions ($Fe^{3+}$) on the octahedral sites (**Figure 1**). Previous experimental measurements show that $Fe_3O_4$ has substantial Fe vacancy concentrations at high oxygen activities and elevated temperatures (900-1400 °C), and the deviation from stoichiometry ($\delta$ in $Fe_{3-\delta}O_4$) increases with $p_{O_2}$.[33] These vacancies are known to form predominantly on the octahedral sublattice and serve as the primary point defects under oxidizing conditions. Through comparison to this known data, we can assess whether our calculations identify octahedral sites as the preferred vacancy location and reproduce experimentally observed vacancy concentrations.

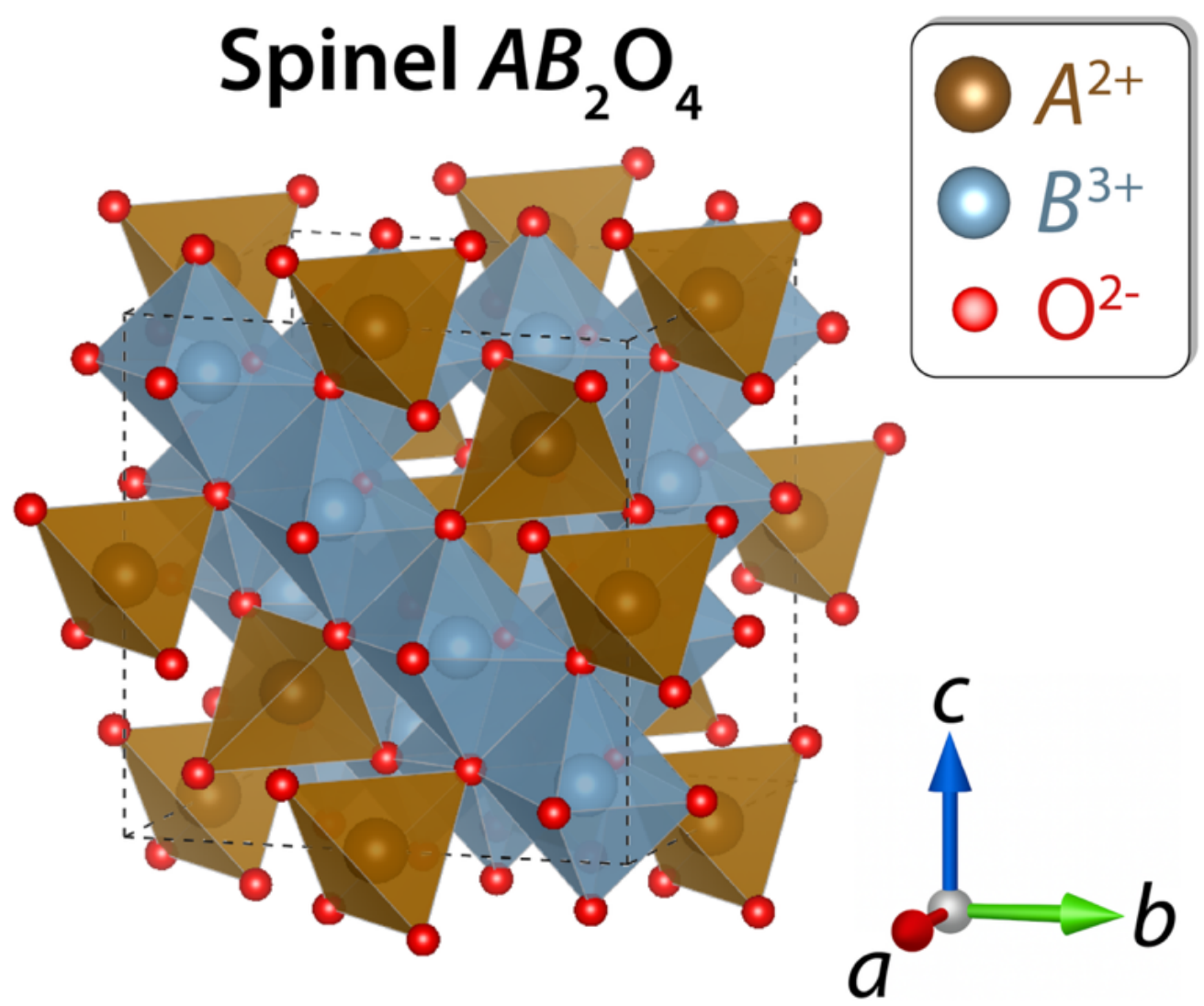

**Figure 1:** Unit cell of the normal spinel crystal structure with general formula $AB_2O_4$. In a typical configuration, $A$ species are divalent ions ($A^{2+}$) that occupy the tetrahedral sites and $B$ species are trivalent ions ($B^{3+}$) that occupy the octahedral sites.

We used a supercell approach[36] to model the defect energetics in $Fe_3O_4$, creating a 112-atom cell with 48 Fe and 64 O atoms. This cell size was found to provide reasonable convergence in the Fe vacancy formation energy (Supplementary Figure 2). Of the 48 Fe sites in our 112-atom cell, 16 occupy tetrahedral sites and 32 occupy octahedral sites. We assumed a ferrimagnetic Fe configuration, initializing the magnetic moments as −4 $\mu_B$ on tetrahedral $Fe^{2+}$ and +3 $\mu_B$ on octahedral $Fe^{3+}$. All calculations employed the PBE+U functional[37,38] with a U-value of 5.3 eV applied to Fe atoms. Further details are provided in **DFT calculations**. Vacancy formation energies were evaluated in the dilute limit by removing one Fe atom from a specific site, relaxing the structure, and computing the energy of the defective cell ($E_{\mathrm{vac}}$) relative to that of $Fe_3O_4$ ($E_{\mathrm{pristine}}$) and Fe ($\mu_{\mathrm{Fe}}$) in its standard state (BCC iron):

$$\Delta E_{\mathrm{f}}\,[\mathrm{V}_{\mathrm{Fe}}''] = E_{\mathrm{vac}} + \mu_{\mathrm{Fe}} - E_{\mathrm{pristine}} \tag{1}$$

In these calculations, the supercell is kept overall charge neutral. The Fe vacancy is therefore presumed to be a doubly charged defect ($\mathrm{V}_{\mathrm{Fe}}''$), with local charge compensation achieved through the oxidation of nearby $Fe^{2+}$ to $Fe^{3+}$ ($2\mathrm{Fe}_{\mathrm{Fe}}^{\bullet} + \mathrm{V}_{\mathrm{Fe}}'' \leftrightarrow 2\mathrm{Fe}_{\mathrm{Fe}}^{\times}$), consistent with the defect mechanism proposed by Warren *et al.* for iron aluminates.[32] This simplified treatment allows us to evaluate the cation vacancy mechanism suggested by experimental observations, without the additional complexity of charged defect equilibria. We acknowledge that a more exhaustive analysis – considering multiple charge states and complexes as performed by Millican *et al.*[34] – could reveal additional complexity in the defect energetics. However, the present approach provides a tractable first assessment of whether the proposed cation vacancy mechanism is thermodynamically plausible under STWS conditions.

The process of computing $\Delta E_{\mathrm{f}}\,[\mathrm{V}_{\mathrm{Fe}}'']$ was repeated for all Fe sites in the 112-atom supercell representing $Fe_3O_4$. Although $Fe_3O_4$ has only two distinct Fe Wyckoff sites, some variation is observed in their computed vacancy energies owing to differences in relaxation pathways, final magnetic configurations, and minor differences in the relaxed cell volumes (ranging from 1218.11 Å$^3$ to 1219.98 Å$^3$). To maintain methodological consistency across $(Fe_{\zeta}Al_{1-\zeta})_3O_4$ compositions, we computed an effective (temperature-weighted) vacancy formation energy using Boltzmann statistics:[39]

$$\exp\left(\frac{-\Delta E_{\mathrm{eff}}[\mathrm{V}_{\mathrm{Fe}}'']}{k_{\mathrm{B}}T}\right) = \frac{1}{N}\sum_i \exp\left(\frac{-\Delta E_{\mathrm{f}}^{i}\,[\mathrm{V}_{\mathrm{Fe}}'']}{k_{\mathrm{B}}T}\right) \tag{2}$$

The summation runs over all computed vacancy formation energies for $N$ total Fe sites, indexed by $i$ and weighted by temperature $T$ in the exponential.

Eqns. 1 and 2 both assume dilute $\mathrm{V}_{\mathrm{Fe}}''$ concentrations, where individual Fe vacancies are sufficiently far apart that they do not interact. At higher concentrations, these vacancies are more likely to occupy nearby sites and interact with one another.[40] To account for such interactions, we evaluated the energetics of 20 randomly sampled configurations containing two or three Fe vacancies within the 112-atom supercell. For each configuration, we computed the formation energy per vacancy as:

$$\Delta E_{\mathrm{f}}(n\mathrm{V}_{\mathrm{Fe}}'')/n = (E_{\mathrm{vac}} + n\mu_{\mathrm{Fe}} - E_{\mathrm{pristine}})/n \tag{3}$$

where $E_{\mathrm{vac}}$ is the energy of the supercell containing $n$ vacancies, $E_{\mathrm{pristine}}$ is the energy of the defect-free supercell, and $\mu_{\mathrm{Fe}}$ is the chemical potential of Fe. For each individual configuration, we define an interaction energy as the difference between its per-vacancy formation energy and the minimum single-vacancy formation energy:

$$\Delta E_{\mathrm{int}}^{i} = \Delta E_{\mathrm{f}}^{i}[n\mathrm{V}_{\mathrm{Fe}}'']/n - \min(\Delta E_{\mathrm{f}}[\mathrm{V}_{\mathrm{Fe}}'']) \tag{4}$$

These interaction energies, shown in Supplementary Figure 3, are positive (indicating repulsive interactions) and generally decrease with increasing vacancy-vacancy separation. To obtain an effective interaction energy for use in thermodynamic modeling, we applied Boltzmann weighting (Eqn. 2) across the 20 sampled configurations to compute an effective multi-vacancy formation energy, $\Delta E_{\mathrm{eff}}[n\mathrm{V}_{\mathrm{Fe}}'']$, and then determined the effective interaction energy per additional vacancy as:

$$\Delta E_{\mathrm{int}} = \Delta E_{\mathrm{eff}}[n\mathrm{V}_{\mathrm{Fe}}'']/n - \Delta E_{\mathrm{eff}}[\mathrm{V}_{\mathrm{Fe}}''] \tag{5}$$

where $\Delta E_{\mathrm{int}} > 0$ indicates repulsive interaction and an energy penalty for forming multiple nearby vacancies.

Equilibrium concentrations of octahedral Fe vacancies in spinels with the general formula $(\mathrm{Fe}_{\zeta}\mathrm{Al}_{1-\zeta})_3\mathrm{O}_4$ were obtained by minimizing the grand potential ($\phi$) with respect to $\delta$ at a given temperature ($T$) and oxygen chemical potential ($\mu_{\mathrm{O}}$):

$$\phi_{(\mathrm{Fe}_{\zeta-\frac{\delta}{3}}Al_{1-\zeta})_3\mathrm{O}_4}(\delta) = \frac{1}{3-\delta}\left[E_{(\mathrm{Fe}_{\zeta-\frac{\delta}{3}}Al_{1-\zeta})_3\mathrm{O}_4}(\delta) - T\big(S_{\mathrm{config}}(\delta) + S_{\mathrm{vib}}(\delta = 0)\big) - \mu_{\mathrm{O}}N_{\mathrm{O}}\right] \tag{6}$$

This potential is normalized by the number of cations per formula unit, $3 - \delta$, while the oxygen count remains fixed at $N_{\mathrm{O}} = 4$. For a chosen cation composition, $\zeta$, a quadratic expression was used to fit $E_{(\mathrm{Fe}_{\zeta-\frac{\delta}{3}}Al_{1-\zeta})_3\mathrm{O}_4}$ as a function of vacancy concentration, based on the $\Delta E_{\mathrm{eff}}$ $[\mathrm{V}_{\mathrm{Fe}}'']$ values obtained from Eqn. 2:

$$E_{(\mathrm{Fe}_{\zeta-\frac{\delta}{3}}Al_{1-\zeta})_3\mathrm{O}_4}(\delta) = E_{(\mathrm{Fe}_{\zeta}Al_{1-\zeta})_3\mathrm{O}_4} + \alpha_1\delta + \alpha_2\delta^2 \tag{7}$$

Where $\alpha_1$ and $\alpha_2$ were fit to match DFT-calculated effective interaction energies in order to obtain a continuous function with respect to $\delta$. The configurational entropy ($S_{\mathrm{config}}$) was estimated by separately assuming ideal solutions on the tetrahedral and octahedral sublattices, noting that vacancies were assumed to form only on the octahedral site:

$$S_{\text{config}}(\delta) = k_B [x_{\text{tet}}^{\text{Fe}} \ln(\text{x}_{\text{tet}}^{\text{Fe}}) + x_{\text{tet}}^{\text{Al}} \ln(x_{\text{tet}}^{\text{Al}})] + 2k_B \left[x_{\text{oct}}^{\text{Fe}} \ln(x_{\text{oct}}^{\text{Fe}}) + x_{\text{oct}}^{\text{Al}} \ln(x_{\text{oct}}^{\text{Al}}) + \frac{\delta}{2} \ln\left(\frac{\delta}{2}\right)\right] \quad (8)$$

Where *x* refers to the fraction of a given site (*tet* for tetrahedral or *oct* for octahedral) occupied by a given species – Fe, Al, or vacancies. The vibrational entropy ($S_{\text{vib}}$) of the pristine spinel structure ($\delta = 0$) with $Fe(_{\zeta}Al_{1-\zeta})_3O_4$ composition was estimated using a structure-based machine learning descriptor from prior work.[41] This descriptor was applied only to the non-defective structures ($\delta = 0$) because it was fit only to stoichiometric crystalline solids. While this term is independent of $\delta$, we include $S_{\text{vib}}(\delta = 0)$ in the espression for $\phi_{(\text{Fe}_{\zeta-\frac{\delta}{3}}Al_{1-\zeta})_3\text{O}_4}(\delta)$ to align the temperature-dependent energy of the solid phases with the temperature-dependent chemical potential of $O_2$. Our assumption is that the difference in vibrational entropies between defective and pristine structures is relatively small. This is supported by Ref. [34], which shows oxygen vacancy formation energies in $FeAl_2O_4$ vary by < 0.1 eV with or without the inclusion of vibrational thermodynamics *via* the quasiharmonic approximation.

The oxygen chemical potential used in our calculations (Eqn. 6) was related to the system temperature and $O_2$ partial pressure according to:

$$\mu_{\text{O}} = \frac{1}{2} G_{\text{O}_2}(T) + \frac{1}{2} RT \ln p_{\text{O}_2} \quad (9)$$

Where $G_{\text{O}_2}(T)$ was taken from experimental thermochemical data.[42] Under reducing conditions, $\mu_{\text{O}}$ is strongly negative (high $T$ and low $p_{\text{O}_2}$), favoring compositions with fewer Fe vacancies because $\phi$ (Eqn. 6) is normalized by the number of cations per formula unit, $3 - \delta$, leading to a larger per-Fe penalty for oxygen-rich states (large $\delta$). Conversely, $\mu_{\text{O}}$ increases under oxidizing conditions (low $T$ and high $p_{\text{O}_2}$), lowering the relative cost of oxygen incorporation. In this regime, the system accommodates the higher oxygen chemical potential by increasing $\delta$, *i.e.*, introducing more Fe vacancies.

The model described above can also be understood *via* the below defect reaction written in Kröger-Vink notation:

$$2\text{Fe}_{\text{Fe}}^{\bullet} + \text{V}_{\text{Fe}}'' + \text{O}_{\text{O}}^{\times} \leftrightarrow 2\text{Fe}_{\text{Fe}}^{\times} + \frac{1}{2}\text{O}_2(\text{g}) \quad (10)$$

Physically, this reaction represents the formation or annihilation of doubly charged Fe vacancies ($\text{V}_{\text{Fe}}''$) in the bulk with charge compensated by the redox of nearby Fe cations ($2\text{Fe}_{\text{Fe}}^{\bullet} + \text{V}_{\text{Fe}}'' \leftrightarrow$ $2\mathrm{Fe}_{\mathrm{Fe}}^{\times}$). Importantly, vacancy formation and annihilation are coupled to surface exchange of Fe. Under reducing conditions, Fe incorporation from the surface is accompanied by oxygen release to the gas phase, whereas under oxidizing conditions Fe removal is coupled to oxygen uptake at the surface. Thus, varying $\mu_{\mathrm{O}}$ directly controls the equilibrium Fe vacancy concentration.

In our analysis, we implicitly assume that octahedral Fe vacancies and their associated $Fe^{2+}/Fe^{3+}$ redox are the dominant defect equilibria under the conditions studied here. This assumption is justified by our calculations showing that alternative defects – including O vacancies, Al vacancies, and interstitials – have substantially higher formation energies that would make their site fractions exceedingly small relative to octahedral Fe vacancies (Supplementary Figure 4). As with our calculation of $\mathrm{V}_{\mathrm{Fe}}''$, these calculations assume charge is compensated by Fe redox, but it is possible that accounting for the formation of other defect complexes or intrinsically charged defects could modify the equilibrium concentrations of $\mathrm{V}_{\mathrm{Fe}}''$ and/or $\mathrm{Fe}_{\mathrm{Fe}}^{\bullet}$ through the laws of mass action (*e.g.*, if Fe sites are oxidized by some other means than $\mathrm{V}_{\mathrm{Fe}}''$ formation). In the scope of this study, we examined the limit where the dominant defect equilibrium is described by Eqn. 10.

**Fe/Al inversion in iron aluminates**

We next examined Fe/Al inversion in the spinel iron aluminate solid-solution system, which spans compositions between magnetite ($Fe_3O_4$) and hercynite ($FeAl_2O_4$) and is bounded by corundum ($Al_2O_3$) at the Al-rich limit. We focus on the composition $(Fe_{\zeta}Al_{1-\zeta})_3O_4$ with $\zeta = 1/3$, corresponding to the $FeAl_2O_4$ endmember. While this composition is commonly referred to as hercynite, it is only stable as a phase-pure compound within a narrow window of temperature and oxygen partial pressure.[43] Outside this stability window, the system typically equilibrates into $Fe_3O_4$-$FeAl_2O_4$ spinel solid solutions in coexistence with corundum. Nevertheless, we adopt single-phase $FeAl_2O_4$ as a model composition to probe inversion and vacancy formation, as it represents the Al-rich limit of the spinel phase field and is the most extensively studied reference composition in this system.

$FeAl_2O_4$ adopts the cubic spinel structure displayed in **Figure 1**, which nominally places all $Fe^{2+}$ on the tetrahedral sublattice and all $Al^{3+}$ on the octahedral sublattice at 0 K. However, neutron diffraction measurements show that substantial inversion (site exchange between Fe and Al) occurs at high temperature.[44] This occurs when the enthalpic penalty for cation inversion is outweighed by the configurational entropy gained.[45] To benchmark our methods against this known behavior, we computed the equilibrium inversion fraction as a function of temperature.

A set of 10 special quasirandom structures (SQSs)[46] were generated in a 112-atom supercell representing $FeAl_2O_4$, each with a varying degree of site inversion. Here, we define the inversion fraction ($x$) as the fraction of tetrahedral sites (initially occupied by Fe) that are now occupied by Al without changing the overall composition. In spinel oxides such as $FeAl_2O_4$, "antisite defects" are sometimes used broadly to describe any deviation from the normal $AB_2O_4$ spinel, including both (1) Fe/Al inversion at fixed composition, which results from paired exchange of Fe and Al between tetrahedral and octahedral sublattices, and (2) composition-driven disorder associated with off-stoichiometry (*e.g.*, $FeAl_2O_4$-$Fe_3O_4$ solid solutions).[34] In this work, we use "inversion" to refer to case (1), meaning the overall Fe:Al ratio is held fixed (consistent with the chosen ζ), and only the site occupancies change as a function of inversion fraction, $x$.

The generated SQSs range from 6.25% inversion (1 of 16 tetrahedral Fe sites exchanged) to 62.5% inversion (10 of 16 exchanged). Quasi-random cation configurations were generated using *sqsgen* (see Supplementary Note 1 for details),[47] with random mixing applied to both the tetrahedral and octahedral sublattices. Each sublattice's pair correlations were independently optimized to match the desired Fe/Al occupancies at the specified inversion fraction. Relaxations were performed using DFT, assuming a ferrimagnetic configuration of antiparallel Fe moments on the tetrahedral and octahedral sites, similar to the setup described previously for $Fe_3O_4$. The resulting energies were used to quantify the enthalpic penalty ($\Delta E_{\mathrm{inv}}$) associated with a specific inversion fraction ($x$) relative to the reference state ($x = 0$):

$$\Delta E_{\mathrm{inv}}(x) = E_{\mathrm{FeAl_2O_4}}(x) - E_{\mathrm{FeAl_2O_4}}(x = 0) \tag{11}$$

These energies are for $FeAl_2O_4$ at fixed composition. While site occupancies vary with $x$, the overall Fe/Al ratio is unchanged. The energy of the normal spinel ($x = 0$) serves as the sole reference against which inverted configuration energies ($x \neq 0$) are compared. A linear fit was constructed from the resulting terms, providing a continuous description of $\Delta E_{\mathrm{inv}}(x)$ from 0 to 62.5% inversion. To determine the equilibrium inversion at finite temperature, we evaluated the Gibbs energy change:

$$\Delta G_{\mathrm{inv}} = \Delta E_{\mathrm{inv}} - TS_{\mathrm{config}} \tag{12}$$

Where $S_{\mathrm{config}}$ is the total ideal solution configurational entropy associated with mixing Fe and Al over the two sublattices. When Fe vacancies are later introduced on the octahedral sublattice, this term generalizes to a three-component mixture (Fe, Al, and $V_{Fe}''$) which further increases $S_{\text{config}}$. For each sampled temperature $T$, the equilibrium inversion fraction was determined by minimizing $\Delta G_{\text{inv}}$ with respect to $x$. The vibrational entropy change associated with inversion was assumed to be zero.

In addition to $FeAl_2O_4$ ($\zeta$ = 1/3), we also used this methodology to study cation inversion in a closely related iron aluminate with increased Al content: $Fe_{1.5}Al_{1.5}O_4$ ($\zeta$ = 1/2). The SQS models were generated for this composition across a similar range of inversion fractions, from which $\Delta E_{\text{inv}}$ was computed using DFT, and $\Delta G_{\text{inv}}$ was obtained using Eqn. 12.

For both iron aluminate compositions ($FeAl_2O_4$ and $Fe_{1.5}Al_{1.5}O_4$), these DFT-relaxed SQSs were also used to calculate Fe vacancy formation energies and concentrations as a function of inversion, following the same procedures outlined for $Fe_3O_4$. All defect calculations employed full atomic relaxations while allowing the cell volume (but not shape) to relax. This relaxation scheme is intended to mimic the experimental situation where high defect concentrations can alter the system volume while preserving the overall symmetry.[30] We focus on cation vacancies as the primary defect to clarify their potential role in STWS, which has been suggested by past experimental work.[32] All vacancy concentrations are evaluated within an equilibrium framework; that is, we assume defect populations re-equilibrate at the cycling temperature and oxygen chemical potential. This is an approximation that neglects kinetic barriers but is supported by prior rapid-heating experiments on iron aluminates showing substantial oxygen release and uptake within seconds at high temperature.[48]

**DFT calculations**

All density functional theory calculations were performed using the projector augmented-wave (PAW) method[37] as implemented in the Vienna Ab Initio Simulation Package (VASP).[27,49] A plane-wave energy cutoff of 520 eV was used with augmentation charges set to 1040 eV. Brillouin zone integrations employed a Γ-centered Monkhorst–Pack grid with a k-point spacing of 0.22 $Å^{-1}$. Electronic convergence was enforced to $10^{-6}$ eV, and ionic relaxations were terminated when forces on all atoms were < 0.01 eV $Å^{-1}$. Spin polarization was included in all calculations, with ferrimagnetic initial states assigned by setting antiparallel moments of −4.0 $\mu_B$ on tetrahedral Fe and +4.0 $\mu_B$ on octahedral Fe sites. Al and O sites were initialized with small 0.1 $\mu_B$ moments. We also tested several randomized spin configurations for pristine and defective $Fe_3O_4$ supercells (Supplementary Figure 5). All were higher in energy by more than 0.02 eV/atom relative to the spin configuration used in this work.

A Hubbard U correction was applied to Fe 3d states using the Dudarev approach (U = 5.3 eV, J = 0.0 eV),[38] consistent with the Materials Project settings.[50] Benchmarking over U = 3.5 to 6.5 eV showed only modest variation in Fe vacancy formation energies (< 0.1 eV, Supplementary Figure 6), possibly owing to partial cancellation of the on-site terms between defective and pristine cells. Although random phase approximation or site-specific U-values obtained from linear-response methods could further improve accuracy, we chose to use a fixed U = 5.3 eV across all Fe sites. Regarding the exchange-correlation functional, we used the Perdew–Burke–Ernzerhof (PBE) generalized gradient approximation (GGA). Fe, Al, and O were described using the standard PAW-PBE pseudopotentials supplied with VASP (no semi-core states). The *pymatgen* package was used to manage calculation results.[51]

## Results

### $Fe_3O_4$ vacancy concentrations

Our DFT calculations on $Fe_3O_4$ indicate $\Delta E_{\mathrm{eff}}\,[\mathrm{V}_{\mathrm{Fe}}'']=1.32$ eV for octahedral Fe vacancies at 1400 °C. This is well below the value of 2.42 eV for tetrahedral vacancies, consistent with prior experimental observations that Fe vacancies strongly prefer the octahedral sublattice.[33] Given this preference, the remaining analysis focuses on octahedral Fe vacancies alone. In **Figure 2a**, we show the distribution of $\Delta E_{\mathrm{f}}\,[\mathrm{V}_{\mathrm{Fe}}'']$ values for configurations with one, two, and three octahedral Fe vacancies per 112-atom cell. Each point corresponds to the energy of a specific vacancy configuration, while stars indicate the effective (temperature-weighted) formation energies at 1400 °C. When $n=1$, $\Delta E_{\mathrm{f}}\,[\mathrm{V}_{\mathrm{Fe}}'']$ ranges from 1.14 to 1.58 eV, agreeing with previous calculations (1.2-1.3 eV).[52] As more vacancies are introduced, repulsive interactions raise the effective formation energy (per vacancy) from 1.41 eV (for $n=2$) to 1.51 eV for ($n=3$). Substituting these values into Eqn. 5 yields a vacancy interaction energy of about 0.09 eV per additional vacancy.

Computed Fe vacancy concentrations (solid curves) are plotted in **Figure 2b** as a function of $p_{\mathrm{O}_2}$ at several temperatures. It should be noted that $Fe_3O_4$ is not thermodynamically stable across the entire $p_{\mathrm{O}_2}$ range shown. At 1400 °C, experimental phase diagrams place the stability window of magnetite roughly between $10^{-9}$ bar (below which reduction to $Fe_{1-x}O$ occurs) and $10^{-1}$ bar (above which oxidation to $Fe_2O_3$ occurs).[53,54] The broader range shown here is intended to illustrate the intrinsic defect response of $Fe_3O_4$ under the assumption that the system is kinetically inhibited from phase separation – consistent with prior reports of Fe vacancy concentrations in $Fe_3O_4$ at 1400 °C.[33] Experimental data from that work are plotted in **Figure 2b** (triangle markers) for comparison.

Our model reproduces the expected trend that more oxidizing conditions favor increased Fe vacancy concentrations. For example, about 3% ($\delta$ = 0.09 in $Fe_{3-\delta}O_4$) of Fe sites are predicted to be vacant at $T = 1200$ °C and $p_{O_2} = 1$ bar. These computed Fe vacancy concentrations also show reasonable agreement with experiment, both in terms of magnitude and onset pressure. There is some noticeable disagreement in the slope of $\delta$ versus $\log_{10}(p_{O_2})$, particularly at high vacancy concentrations (> 1%). Our interaction term (Eqn. 5) is based on a finite sampling of defect configurations and may not fully represent longer-range or higher-order effects. Because the slope of δ versus $p_{O_2}$ is especially sensitive to these interactions (as later shown in **Figure 7**), small inaccuracies in $\Delta E_{\mathrm{int}}$ can lead to substantial deviations from experiment at large $\delta$. Nevertheless, we believe there is sufficient agreement with experiment to justify the application of these methods to more complex spinel compositions.

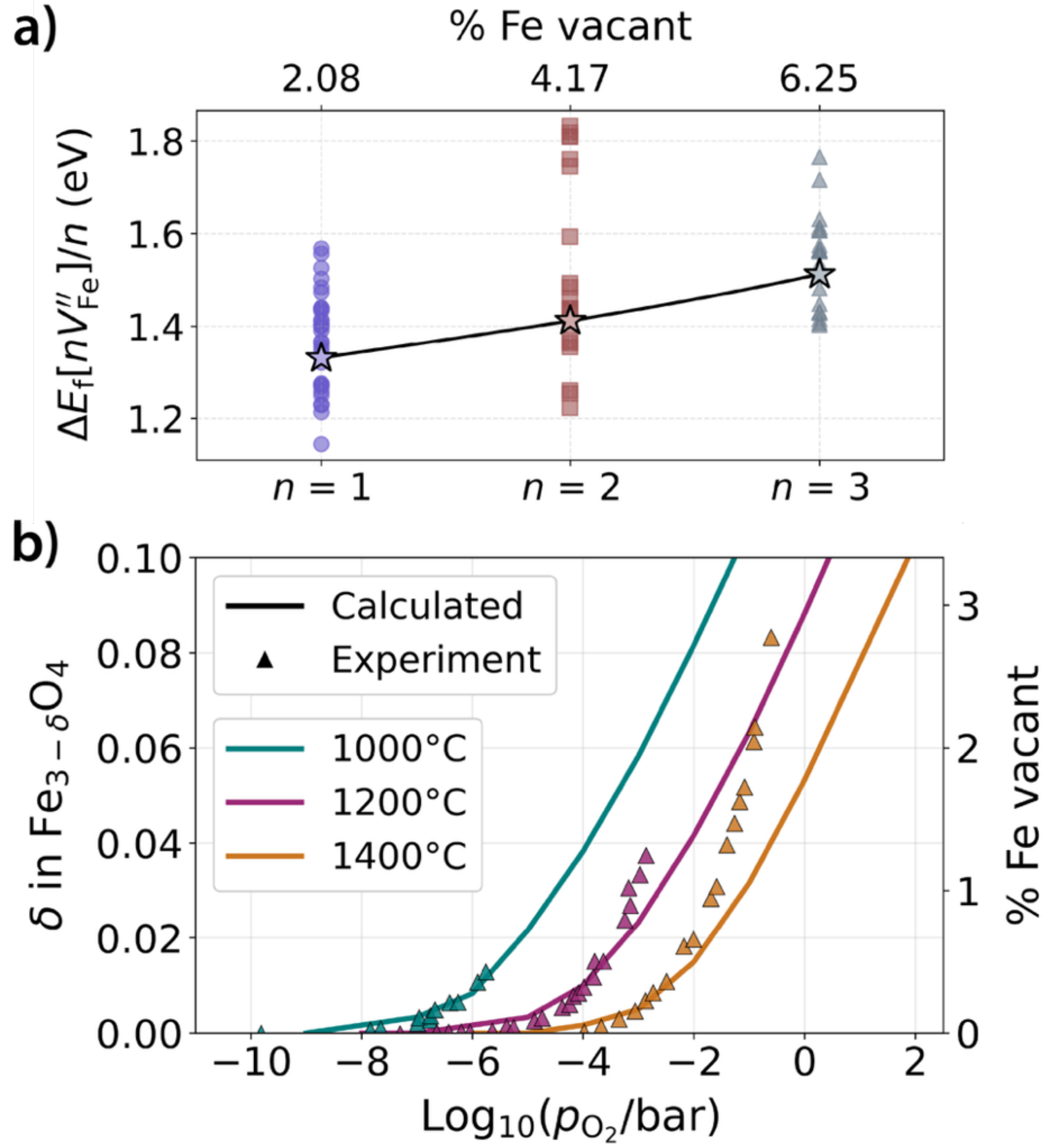

**Figure 2:** (a) Octahedral Fe vacancy formation energies in $Fe_3O_4$ computed using PBE+U. Each point corresponds to a single vacancy configuration. Stars represent effective vacancy formation energies computed at 1400 °C. The lower $x$-axis ($n$) indicates the number of Fe vacancies in a 112-atom cell while the upper $x$-axis describes the % concentration of these vacancies. (b) Equilibrium amounts of octahedral Fe vacancies per $Fe_3O_4$ formula unit ($\delta$, left $y$-axis) and concentrations (%, right $y$-axis) determined using the calculated formation energies and plotted with respect to partial pressure of oxygen ($p_{O_2}$ in bar) on a log scale. For comparison, experimentally determined Fe vacancy concentrations[33] are plotted as triangle markers.

### Fe/Al inversion thermodynamics

Before analyzing Fe vacancies in $(Fe_{\zeta}Al_{1-\zeta})_3O_4$ spinels, we first assessed their equilibrium cation (Fe and Al) configurations since inversion is likely at the high temperatures relevant to STWS. DFT-computed energies (black points) and temperature-dependent Gibbs energy curves of partially inverted $FeAl_2O_4$ ($\zeta$ = 1/3) configurations are shown in **Figure 3a**. At 0 K, all configurations incur a positive enthalpic penalty ($\Delta E_{\mathrm{inv}} > 0$), reflecting the higher internal energy of structures with antisite defects. However, as temperature increases, configurational entropy stabilizes the partially inverted structures, producing a minimum in the Gibbs energy ($\Delta G_{\mathrm{inv}}$) at non-zero inversion fraction. The predicted equilibrium inversion fraction is plotted as a function of temperature in **Figure 3b**, alongside experimental data from neutron diffraction measurements.[44] The calculated inversion rises steadily with temperature, reaching 32% at 1400 °C. This trend agrees reasonably well with experimental measurements, with some (~5%) overestimation at high temperatures. Previous CALPHAD models have also predicted higher inversion fractions than experiment,[55] consistent with our DFT calculations.

We extended a similar analysis to $Fe_{1.5}Al_{1.5}O_4$ ($\zeta$ = 1/2), which adopts a cation arrangement where $Fe^{2+/3+}$ partially occupies both the tetrahedral and octahedral sublattices even in the absence of inversion, while $Al^{3+}$ fills the remaining octahedral sites. Unlike $FeAl_2O_4$, whose 0 K ground state (the normal spinel) places all Fe on tetrahedral sites and all Al on octahedral sites, $Fe_{1.5}Al_{1.5}O_4$ inherently contains 25% Fe occupancy on the octahedral sublattice in its 0 K ground state. This represents a baseline level of cation disorder that arises from the composition itself rather than thermal activation. We note that this is distinct from the antisite disorder associated with $FeAl_2O_4$-$Fe_3O_4$ solid solutions, which were examined by Millican *et al.*,[34] where the Fe/Al ratio varies. Here, we maintain the fixed $Fe_{1.5}Al_{1.5}O_4$ stoichiometry and examine how additional site exchange (inversion) beyond the baseline configuration affects the system thermodynamics. We computed the temperature-dependent Gibbs energy change associated with this additional inversion and determined a high equilibrium inversion fraction of 41% at 1400 °C (**Table 1**). This increase relative to $FeAl_2O_4$ may be attributed to the fact that $Fe_{1.5}Al_{1.5}O_4$ already exhibits partial cation disorder at 0 K, likely reducing the enthalpic penalty for additional site inversion.

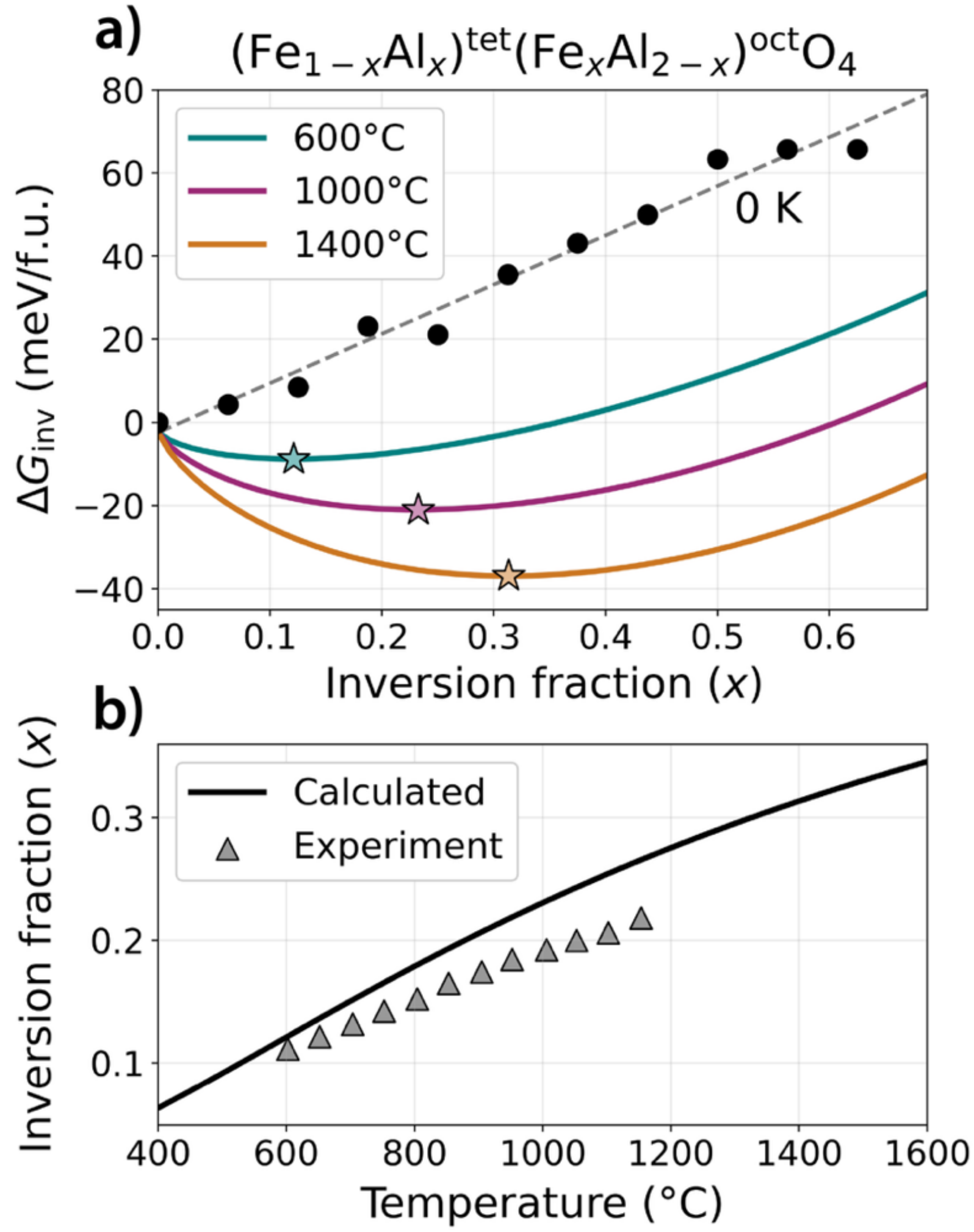

**Figure 3:** (a) Gibbs energy change ($\Delta G_{\text{inv}}$) of $FeAl_2O_4$ with respect to Fe/Al inversion ($x$). The 0 K ground state of $FeAl_2O_4$ places all $Fe^{2+}$ in tetrahedral sites and all $Al^{3+}$ in octahedral sites. Exchange between these sites incurs a penalty on the system enthalpy, as shown by the black dots in the top panel representing DFT-computed energies of partially inverted configurations. However, the curves show that some degree of inversion (equilibrium $x$ denoted by stars) becomes favorable at finite temperature owing to increased configurational entropy. (b) These equilibrium inversion fractions ($x$) are plotted as a function of temperature, along with experimental values (triangles) determined using neutron diffraction measurements.[44]

**Table 1:** Composition, structure formula, equilibrium inversion fraction ($x$), and total Fe content in tetrahedral and octahedral sites of three Fe-Al spinel compounds at 1400°C. The inversion fraction is defined as the percentage of tetrahedral sites occupied by Al. These values were derived from DFT calculations based on the PBE+U functional. Random cation mixing on each sublattice was assumed to estimate configurational entropy.

| Composition | Structure formula | $x$ at 1400 °C | Total $Fe^{tet}$ | Total $Fe^{oct}$ |
|---|---|---|---|---|
| $Fe_3O_4$ | $(Fe)^{tet}(Fe_2)^{oct}O_4$ | N/A | 1.00 | 2.00 |
| $Fe_{1.5}Al_{1.5}O_4$ | $(Fe_{1-x}Al_x)^{tet}(Fe_{0.5+x}Al_{1.5-x})^{oct}O_4$ | 0.41 | 0.59 | 0.91 |
| $FeAl_2O_4$ | $(Fe_{1-x}Al_x)^{tet}(Fe_xAl_{2-x})^{oct}O_4$ | 0.32 | 0.68 | 0.32 |

**Fe vacancy energetics with inversion**

Having established that cation inversion is thermodynamically favored in $(Fe_\zeta Al_{1-\zeta})_3O_4$ spinels at high temperature, we next examined how such inversion affects the formation of octahedral Fe vacancies. To isolate this effect, we analyzed a series of partially inverted SQS configurations spanning a range of inversion fractions. For each SQS, we identified all unique octahedral Fe sites with distinct local coordination environments and constructed a separate defective structure by removing Fe from that site. Only octahedral vacancies are presented, as their effective formation energies $\Delta E_{\mathrm{eff}}[\mathrm{V}_{\mathrm{Fe}}'']$ (defined in Eqn. 2) are consistently at least 1 eV lower than that of tetrahedral Fe vacancies. In **Figure 4**, we show $\Delta E_{\mathrm{eff}}[\mathrm{V}_{\mathrm{Fe}}'']$ for octahedral Fe vacancies in $FeAl_2O_4$ ($\zeta$ = 1/3, green circles) and $Fe_{1.5}Al_{1.5}O_4$ ($\zeta$ = 1/2, orange squares), plotted as a function of their inversion fraction. Dashed lines represent linear fits to the computed values, and for $Fe_3O_4$ ($\zeta$ = 1) – where inversion is not defined since it contains no Al – a single horizontal dashed line shows the effective vacancy formation energy. All data correspond to dilute vacancies ($n = 1$ vacancy per 112-atom cell) modeled at 1400 °C.

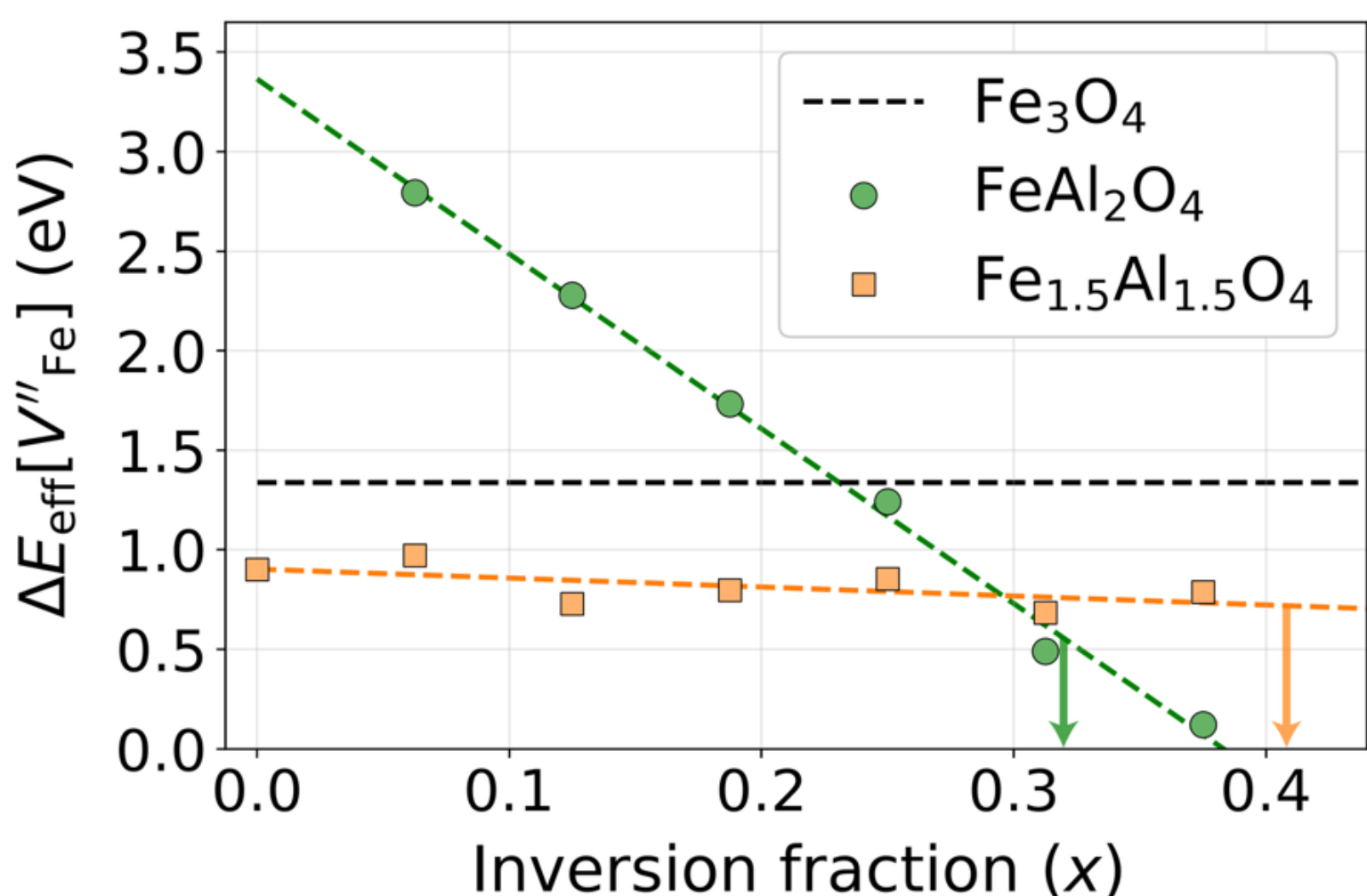

**Figure 4:** Effective formation energies of octahedral Fe vacancies at 1400 °C in iron aluminates, evaluated with varying degrees of Fe/Al inversion ($x$). For $FeAl_2O_4$ (green circles) and $Fe_{1.5}Al_{1.5}O_4$ (orange squares), markers represent DFT-calculated effective vacancy formation energies and dashed lines show linear fits to these values. For $Fe_3O_4$, where inversion cannot occur due to the absence of Al, a single horizontal dashed line indicates the constant vacancy formation energy. This plot only includes dilute single-vacancy formation energies ($n = 1$). Green and orange arrows indicate equilibrium inversion fractions at 1400 °C.

We observe a strong dependence of $\Delta E_{\text{eff}}[V''_{\text{Fe}}]$ on the inversion fraction in $FeAl_2O_4$, consistent with prior results.[45] At low inversion (6.25%), octahedral Fe vacancies have a large (unfavorable) formation energy of 2.79 eV. However, this drops dramatically to just 0.12 eV at 37.5% inversion. Linear interpolation predicts a formation energy of 0.62 eV at 32% inversion – the equilibrium value at 1400 °C – suggesting that octahedral Fe vacancies should be reasonably accessible under conditions relevant to STWS. While negative formation energies are predicted for inversion levels above 37.5%, such cases are likely unphysical and would not necessarily lead to higher Fe vacancy concentrations. The high temperatures needed to drive such inversion would impose strongly reducing conditions that lower oxygen content in the structure, decreasing the O:Fe ratio and thereby suppressing Fe vacancies. The concentration of these vacancies would also be limited by repulsive interactions between them.

Inversion likely facilitates Fe vacancy formation by placing the system in a higher-enthalpy reference state. Inverted Fe atoms occupy the octahedral sites primarily to increase configurational entropy, despite being enthalpically unfavorable at 0 K (**Figure 3a**). Because these inverted sites are already high in energy, removing an Fe atom from them imposes only a modest additional penalty, while further increasing entropy by creating a vacancy through the introduction of a third constituent on the octahedral sublattice (Fe/Al/$V_{Fe}''$). As a result, the net cost for vacancy formation is substantially diminished in these inverted configurations.

To assess whether octahedral Fe vacancies are indeed the dominant redox-active defects under STWS conditions, we computed formation energies for several alternative defect types in $FeAl_2O_4$, including tetrahedral Fe vacancies, Al vacancies, O vacancies, and Fe/Al interstitials. Distributions of defect formation energies are shown in Supplementary Figure 4. In the normal spinel, oxygen vacancies have the lowest formation energies (3.6-3.7 eV at $T = 1400$ °C and $p_{O_2} = 10^{-5}$ bar), consistent with the findings of Millican *et al.*[34] All other computed defects have formation energies > 4 eV. With Fe/Al inversion, low-energy configurations become accessible largely due to broadening of the defect formation energy distributions. In particular, oxygen vacancies and tetrahedral Fe vacancies become more competitive, with formation energies of each as low as 3.2 eV. This further supports past findings that oxygen vacancies pair favorably with Fe/Al antisite defects.[34] However, none of these point defects approach the low formation energies (< 1 eV) displayed by octahedral Fe vacancies under the same conditions, supporting the importance of these defects in partially inverted iron aluminates.

In contrast to $FeAl_2O_4$, we find that $Fe_{1.5}Al_{1.5}O_4$ exhibits a weaker dependence on inversion. It begins with a relatively low vacancy formation energy of 0.90 eV at 0% inversion – well below that of $FeAl_2O_4$ at similar inversion levels. However, this formation energy decreases only slightly with site inversion, making Fe vacancies less favorable in $Fe_{1.5}Al_{1.5}O_4$ than in $FeAl_2O_4$ at inversion levels > 30%. Recall from **Table 1** that an equilibrium inversion fraction of 41% was predicted for $Fe_{1.5}Al_{1.5}O_4$ under conditions relevant for STWS. In this configuration, $\Delta E_{\mathrm{eff}}[V_{Fe}'']$ is about 0.72 eV – still accessible, but higher than the value predicted for $FeAl_2O_4$ under the same conditions.

The weaker dependence on inversion observed for $Fe_{1.5}Al_{1.5}O_4$ likely stems from its nominal cation arrangement, which already includes 25% Fe occupancy on the octahedral sublattice in its 0 K ground state. Because the system begins with octahedral Fe occupation, inversion causes only a modest change in the reference energy ($E_{\mathrm{pristine}}$ in Eqn. 1), resulting in a limited effect on the vacancy formation energy. To illustrate this point, we compare the two limiting cases: 1) $Fe_3O_4$ has all octahedral sites occupied by Fe, and its dilute Fe vacancy formation energy is relatively high at 1.34 eV; 2) $FeAl_2O_4$ contains no octahedral Fe at 0 K, but as inversion places Fe on these sites, the vacancy formation energy drops substantially to 0.62 eV at 1400 °C. This comparison suggests that $\Delta E_{\mathrm{eff}}[\mathrm{V}_{\mathrm{Fe}}'']$ is not solely determined by the fraction of Fe on the octahedral sites, but also by the relative site energy associated with octahedral Fe. Only when octahedral Fe is enthalpically unfavorable, as in $FeAl_2O_4$, does inversion significantly decrease the cost of forming cation vacancies.

In **Figure 4**, only dilute vacancy concentrations ($n = 1$ vacancy per 112-atom cell) are considered. To assess the impact of vacancy-vacancy interactions at higher concentrations, we performed additional calculations using configurations with $n = 2$ and 3 vacancies per 112-atom supercell, following the methodology described earlier (see **$Fe_3O_4$ vacancy calculations**). The results are shown in **Figure 5**, where effective vacancy formation energies $\Delta E_{\mathrm{eff}}[\mathrm{V}_{\mathrm{Fe}}'']$ for three $(Fe_{\zeta}Al_{1-\zeta})_3O_4$ spinels with $\zeta$ = 1/3 ($FeAl_2O_4$), 1/2 ($Fe_{1.5}Al_{1.5}O_4$), and 1 ($Fe_3O_4$) are shown. These formation energies were computed for configurations with equilibrium inversion at 1400 °C, then normalized per vacancy and grouped by defect number ($n$).

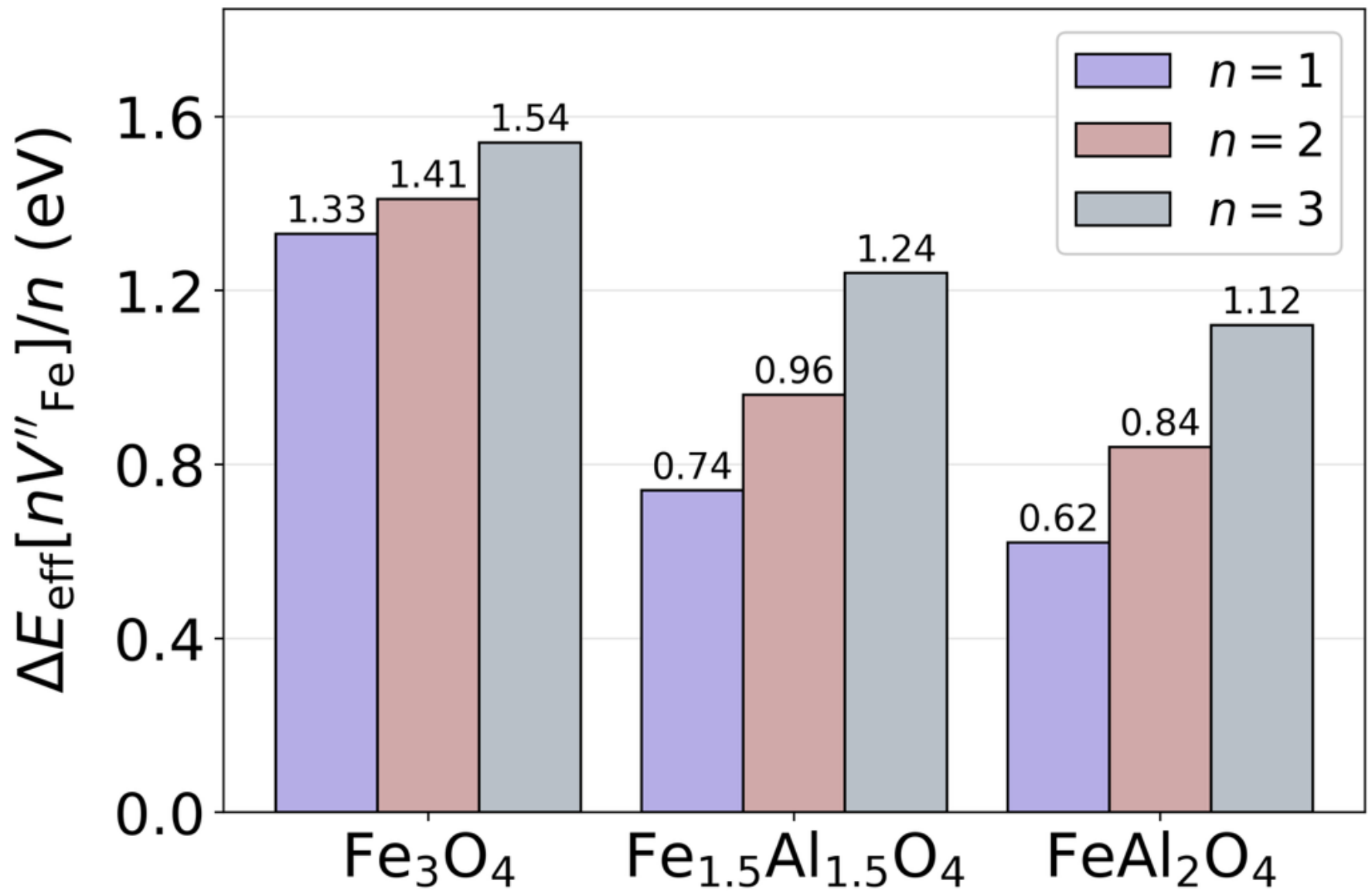

**Figure 5:** Effective octahedral Fe vacancy formation energies in Fe-Al spinel compounds. These are computed using PBE+U on structures with equilibrium Fe/Al inversion at 1400 °C. This corresponds to 41% inversion for $Fe_{1.5}Al_{1.5}O_4$ and 32% inversion for $FeAl_2O_4$. The bar color in the plot is related to the number of Fe vacancies ($n$) in each 112-atom cell. All formation energies are normalized to a single-vacancy basis.

Two trends emerge from the results in **Figure 5**. First, the dilute ($n = 1$) vacancy formation energies follow the order $Fe_3O_4$ > $Fe_{1.5}Al_{1.5}O_4$ > $FeAl_2O_4$. Second, the vacancy-vacancy interaction energies differ substantially across systems. $Fe_3O_4$ exhibits the weakest interactions, with an average $\Delta E_{\mathrm{int}}\,[\mathrm{V}''_{\mathrm{Fe}}]$ of only about 0.09 eV per additional vacancy formed. In contrast, $Fe_{1.5}Al_{1.5}O_4$ and $FeAl_2O_4$ both show much stronger vacancy interaction energies ranging from 0.22 to 0.25 eV. Therefore, while these iron aluminates exhibit lower dilute vacancy formation energies than $Fe_3O_4$, their vacancy concentrations may be somewhat limited by these strong (repulsive) interactions. In other words, increasing Al content (decreasing ζ) makes initial Fe vacancy formation more favorable, but it also strengthens vacancy-vacancy interactions, limiting the Fe vacancy concentration.

**STWS hydrogen yield**

The ability to form and consume cation vacancies is directly linked to $H_2$ yield in STWS, as described by the below equation adapted from previous work:[32]

$$\left(\frac{3-\delta_f}{4}\right)\left(\mathrm{Fe}_\zeta\mathrm{Al}_{1-\zeta}\right)_{3-\delta_i}\mathrm{O}_4 + (\Delta\delta)\mathrm{H_2O} \leftrightarrow \left(\frac{3-\delta_i}{4}\right)\left(\mathrm{Fe}_\zeta\mathrm{Al}_{1-\zeta}\right)_{3-\delta_f}\mathrm{O}_4 + (\Delta\delta)\mathrm{H}_2 \quad (13)$$

$\Delta\delta$ denotes the change in cation vacancy concentration ($\delta_f - \delta_i$) during a redox cycle and determines the equilibrium $H_2$ yield per cycle. Such a change can be induced by varying either the temperature or the oxygen chemical potential. We focus here on the latter, which is more relevant to experimental STWS cycles that operate isothermally at 1400 °C. To be consistent with prior studies,[32] where $p_{\mathrm{O}_2}$ was defined by typical $H_2O$ input pressures and their equilibrium thermolysis at 1400 °C, all $H_2$ yields in this work were computed over $10^{-5}$ bar $\leq p_{\mathrm{O}_2} \leq 3.84 \times 10^{-4}$ bar.

Equilibrium Fe vacancy concentrations are plotted as a function of $p_{\mathrm{O}_2}$ in **Figure 6a** for three $(\mathrm{Fe}_\zeta\mathrm{Al}_{1-\zeta})_3\mathrm{O}_4$ spinels: $\zeta$ = 1/3 ($FeAl_2O_4$), 1/2 ($Fe_{1.5}Al_{1.5}O_4$), and 1 ($Fe_3O_4$). These concentrations were determined by minimizing the grand potential (Eqn. 6) of each system in its equilibrium inversion state at 1400 °C. Of these three compositions, $FeAl_2O_4$ accommodates the highest concentration of cation vacancies, with $\delta$ ranging from 0.179 to 0.221 over $10^{-5}$ bar $\leq p_{\mathrm{O}_2} \leq 3.84 \times 10^{-4}$ bar. $Fe_{1.5}Al_{1.5}O_4$ exhibits somewhat lower concentrations ($\delta$ = 0.080 to 0.106), reflecting its higher vacancy formation energy (0.72 eV) at equilibrium inversion. In contrast to both aluminates, $Fe_3O_4$ supports only modest cation vacancy concentrations ($\delta < 0.1$) as it has an even higher vacancy formation energy (1.32 eV).

These trends in Fe vacancy concentrations and their response to changes in $O_2$ pressure have a direct influence on the hydrogen yield of the iron aluminates. As shown in **Figure 6b**, our calculations indicate that $FeAl_2O_4$ outperforms $Fe_{1.5}Al_{1.5}O_4$ and $Fe_3O_4$ for STWS, achieving a predicted $H_2$ yield of 361 μmol/g. This result confirms that cation vacancies alone can facilitate substantial water splitting, a conclusion that echoes recent experimental findings. However, we generally under-predict the $H_2$ yield compared to prior reports,[32] which exceed 450 μmol/g over a comparable $p_{\mathrm{O}_2}$ range. To better understand this discrepancy, we compare the calculated (lines) and experimental (markers) vacancy concentrations in **Figure 6a**. While the magnitudes of these concentrations agree reasonably well, there is a noticeable mismatch between experiment and theory when it comes to the slope of $\delta$ versus $\log_{10}(p_{\mathrm{O}_2})$ – a feature that is closely tied to vacancy-vacancy interactions.[56] In addition to differences in these interactions, it is possible that some mismatch arises because our calculations consider only single-phase spinels, whereas experiments may also involve two-phase equilibria (*e.g.*, with $Al_2O_3$).[43]

The limited water-splitting capacity of $Fe_3O_4$, despite its ability to form Fe vacancies, can also be understood from these results. While octahedral Fe vacancies do form in magnetite, the equilibrium concentrations and their response to changes in oxygen partial pressure are too modest to drive substantial $H_2$ production. This is a direct consequence of the relatively high vacancy formation energy (1.32 eV) in $Fe_3O_4$, which limits both the absolute vacancy concentration and the magnitude of $\Delta\delta$ achievable over a given $p_{O_2}$ swing. In contrast, the Al-rich spinels benefit from cation inversion that dramatically lowers the effective cost of forming octahedral Fe vacancies – *e.g.*, down to 0.62 eV in $FeAl_2O_4$ under equilibrium inversion at 1400 °C. This lower formation energy enables larger swings in vacancy concentration across the redox cycle and correspondingly higher $H_2$ yields, underscoring the critical role of antisite disorder (inversion) in unlocking the water-splitting potential of iron aluminate spinels.

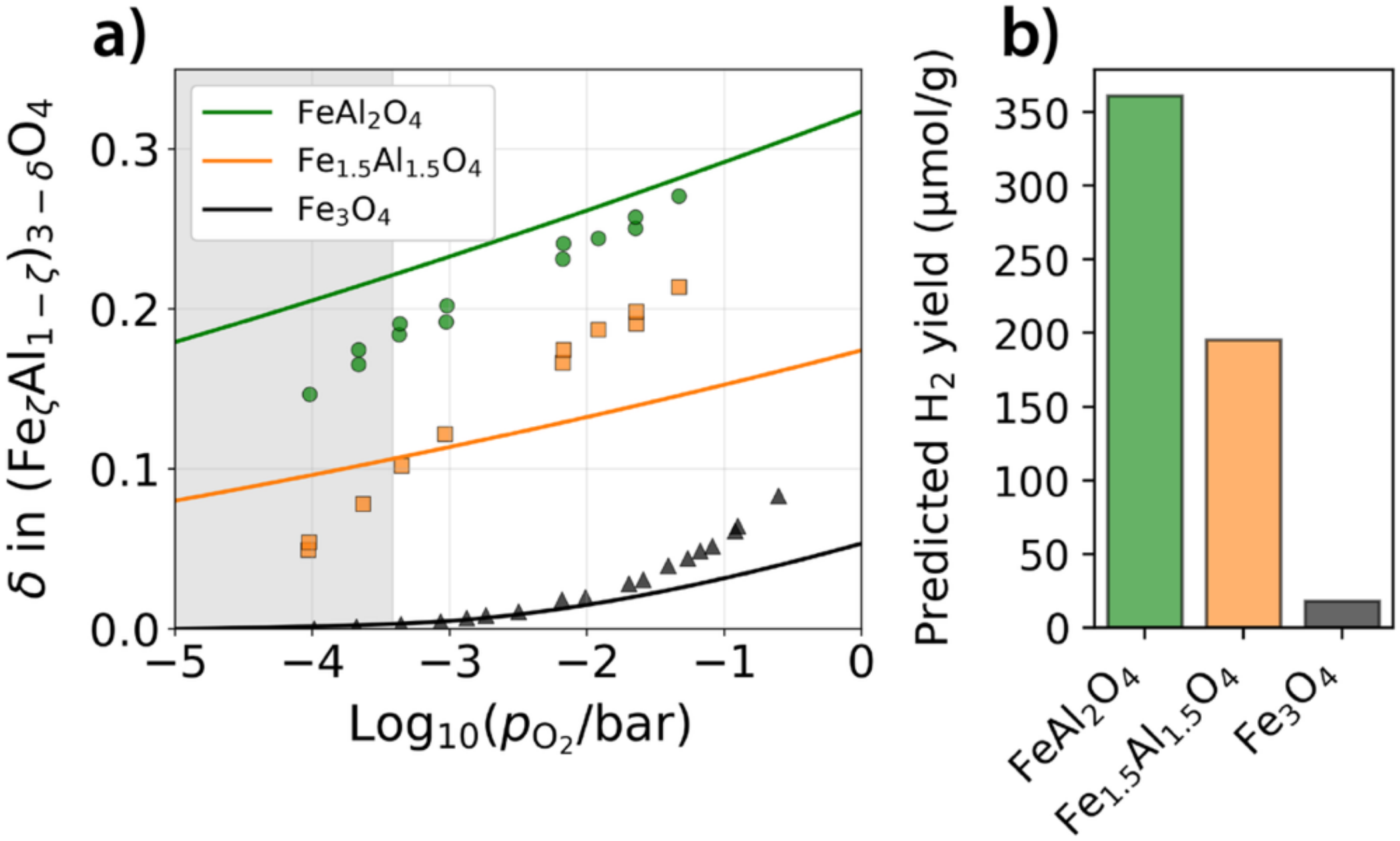

**Figure 6:** (a) Cation vacancy concentrations versus $O_2$ partial pressure. These concentrations were computed from the equilibrium structure (with Fe/Al inversion) of each composition at 1400 °C. The solid lines represent calculated data, while the markers represent experimental data from prior work.[32] (b) $H_2$ yields derived from the calculated vacancy concentrations, assuming a pressure swing ($\Delta p_{O_2}$) between $10^{-5}$ bar and $3.84 \times 10^{-4}$ bar at 1400 °C (shaded region in panel a).

To better understand the calculated $H_2$ production in these iron aluminates, we systematically examined how two key parameters affect vacancy concentration and yield: the effective Fe vacancy formation energy, $\Delta E_{\mathrm{eff}}\,[\mathrm{V}_{\mathrm{Fe}}'']$, and the vacancy-vacancy interaction strength, $\Delta E_{\mathrm{int}}\,[\mathrm{V}_{\mathrm{Fe}}'']$. The molar mass of the active material also plays some role, since the yield is normalized per gram, but is less critical than the underlying defect thermodynamics. In **Figures 7a** and **7c**, we illustrate how variations in the vacancy formation and interaction energy affect the equilibrium $\delta$ and resulting $H_2$ yield in $FeAl_2O_4$ by varying $\Delta E_{\mathrm{eff}}\,[\mathrm{V}_{\mathrm{Fe}}'']$ while holding $\Delta E_{\mathrm{int}}\,[\mathrm{V}_{\mathrm{Fe}}'']$ constant. These calculations do not pertain to any particular material but rather serve to disentangle the effects of each parameter within our model for a cation vacancy-driven redox cycle. Increasing $\Delta E_{\mathrm{eff}}\,[\mathrm{V}_{\mathrm{Fe}}'']$ systematically reduces the overall vacancy concentration but has a relatively weak effect on the slope of each $\delta$ curve. Large changes to $\Delta E_{\mathrm{eff}}\,[\mathrm{V}_{\mathrm{Fe}}'']$ are therefore needed to influence the $H_2$ yield (**Figure 7c**), which correlates directly with the change in $\delta$ as $p_{\mathrm{O_2}}$ is varied.

The converse situation is explored in **Figures 7b** and **7d**, where $\Delta E_{\mathrm{int}}\,[\mathrm{V}_{\mathrm{Fe}}'']$ is varied while $\Delta E_{\mathrm{eff}}\,[\mathrm{V}_{\mathrm{Fe}}'']$ is held fixed. In this case, we observe a more pronounced effect on the slope of each $\delta$ curve, which translates into substantial differences in $H_2$ yield (**Figures 7d**). Stronger repulsive interactions flatten the $\delta$ curve, meaning there is less change in the vacancy concentration with $p_{\mathrm{O_2}}$, and this translates to lower $H_2$ yields. From these trends, several design principles emerge. To maximize $H_2$ yield in materials that operate through a cation-vacancy mechanism, materials should exhibit: 1) low vacancy formation energy ($< 1$ eV), ensuring that sufficient defect concentrations can form under STWS conditions; and 2) weak vacancy-vacancy interactions ($< 0.3$ eV), allowing the system to maximize the change in vacancy concentration ($\Delta\delta$) by responding sharply to changes in oxygen chemical potential.

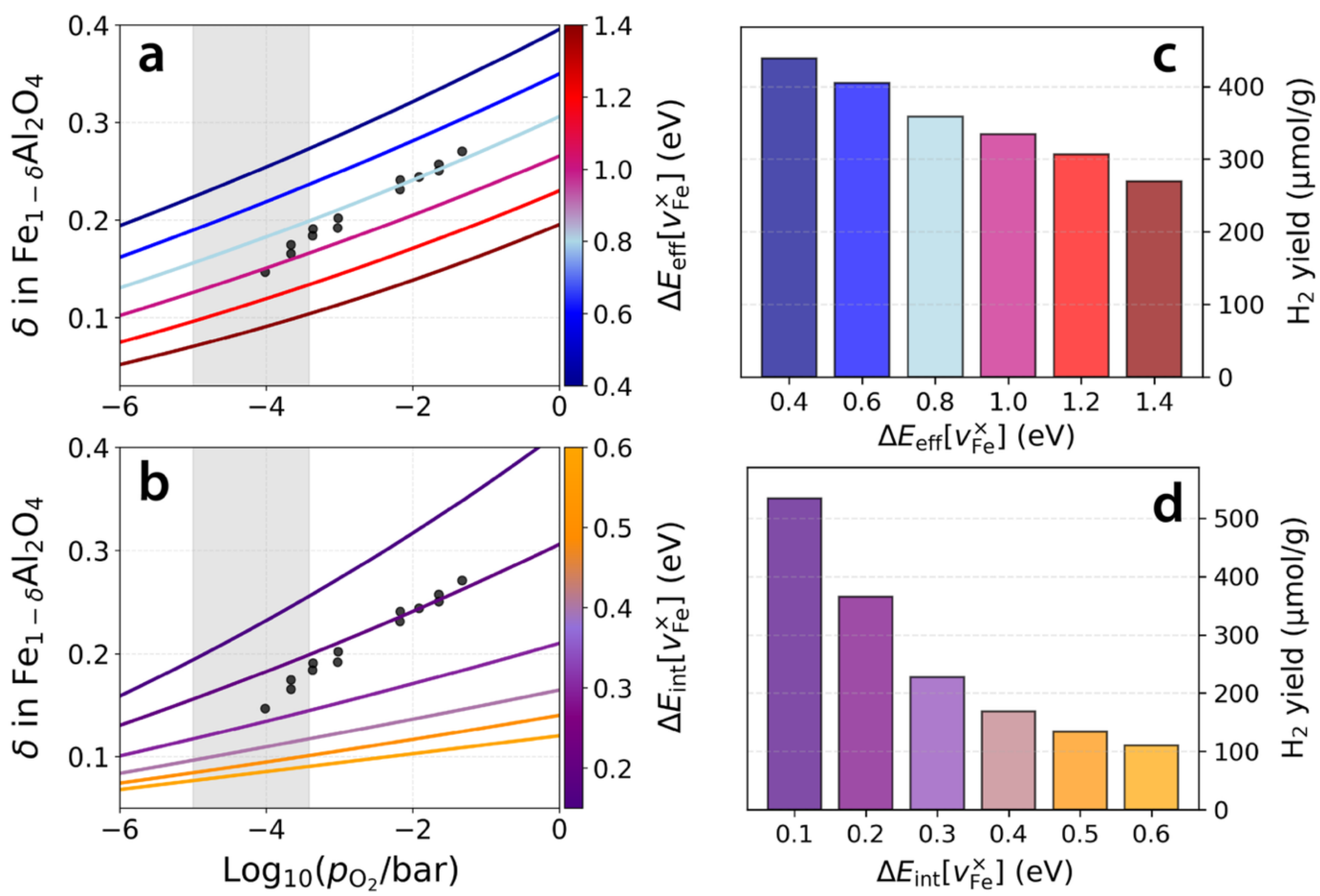

**Figure 7:** (a, b) Equilibrium concentrations of octahedral Fe vacancies in $FeAl_2O_4$, simulated with artificially varied (a) effective formation energies and (b) interaction energies. When changing the vacancy formation energy, a fixed interaction energy of $\Delta E_{\text{int}}\,[\mathrm{V}_{\mathrm{Fe}}''] = 0.2$ eV is used. Conversely, when changing the interaction energy, a constant formation energy of $\Delta E_{\text{eff}}\,[\mathrm{V}_{\mathrm{Fe}}''] = 0.8$ eV is used. Curves represent calculated values based on these hypothetical formation and interaction energies. Black dots correspond to experimental data.[32] (c, d) $H_2$ yields derived from the calculated data, assuming $p_{\mathrm{O_2}}$ between $10^{-5}$ bar and $3.84 \times 10^{-4}$ bar at 1400 °C (shaded regions in panels a and b).

## Discussion

Solar thermochemical water splitting (STWS) for hydrogen production has traditionally relied on oxygen vacancy formation as the dominant redox mechanism. This is well-established in fluorite-structured materials such as $CeO_2$, where reversible oxygen release and uptake drive $H_2$ generation at high temperatures. In contrast, the feasibility of STWS *via* cation vacancies has remained more contentious. While experimental studies[30,32] have suggested that cation deficiency plays a key role in the redox behavior of iron aluminate spinels, theoretical support for this mechanism has so far been limited.

In this work, we examined the feasibility of a cation-vacancy-mediated STWS cycle in iron aluminate spinels of the form $(Fe_{\zeta}Al_{1-\zeta})_3O_4$ using first-principles thermodynamics to evaluate equilibrium defect concentrations and $H_2$ yields. Our calculations predict that up to 15% of cation sites can accommodate Fe vacancies under typical STWS conditions (1400 °C). When coupled to changes in the oxygen chemical potential across a realistic $O_2$ pressure swing ($10^{-5}$ bar to $3.84 \times 10^{-4}$ bar), these vacancies can drive substantial $H_2$ production. While our calculations underpredict the total $H_2$ yield observed experimentally – *e.g.*, 361 versus 450 μmol/g for $FeAl_2O_4$ – they show that Fe vacancies alone still account for a substantial fraction of the measured capacity.

Previous theoretical work on $FeAl_2O_4$ has primarily examined oxygen-vacancy-driven redox. Millican *et al.*[34] showed that oxygen vacancies paired with Fe/Al antisite defects are the most favorable defect complexes in Fe-rich compositions, with predicted oxygen deficiencies up to $\delta = 2 \times 10^{-3}$ at 1400 °C and $p_{O_2} = 10^{-4}$ atm. However, this remains well below what is needed to fully account for the $H_2$ yields observed experimentally.[30,32] In light of these findings, our study focused on Fe vacancies and how their formation energies evolve with Fe/Al inversion at fixed composition (*i.e.*, without changing the overall Fe/Al ratio). We find that while cation vacancies are not favorable under low-inversion conditions, this changes under typical STWS conditions, where a substantial fraction of cation sites become inverted. In this regime, Fe vacancy formation energies decrease from > 3 eV to 0.62 eV, making such defects thermodynamically accessible. The resulting Fe vacancy concentrations ($\delta \sim 0.2$) far exceed those of oxygen vacancies, enabling higher $H_2$ yields. That said, these two mechanisms – oxygen vacancies in Fe-rich compositions and Fe vacancies facilitated by inversion – are not mutually exclusive and may both contribute to the redox behavior of iron aluminates.

There are several simplifications in our approach that merit discussion. First, we treat Fe vacancies exclusively as doubly charged defects with local charge compensation *via* $Fe^{2+}/Fe^{3+}$ redox, following the mechanism proposed by Warren *et al.*[32] A comprehensive charged defect analysis would provide a more complete picture, but the reasonable agreement between our predictions and experimental data suggests that the doubly-charged vacancies with compensation by neighboring Fe oxidation capture the dominant behavior. However, we cannot rule out contributions from alternative charge states or defect complexes, such as those studied previously.[34] Second, although we include an effective vacancy-vacancy interaction term by sampling multi-vacancy configurations, we do not explicitly treat defect association equilibria (*e.g.*, bound vacancy-antisite complexes) or the corresponding entropy penalties that may result from short-range ordering, as reported by Millican *et al.*[34] Frenkel defects comprised of Fe vacancy-interstitial pairs may also play a role. These were not assessed in this work, but the isolated interstitial defects we did evaluate in $FeAl_2O_4$ had very high formation energies (> 5 eV), much higher than the interstitial formation energies reported previously for $Fe_3O_4$.[57]

Beyond validating the feasibility of a cation vacancy mechanism, our results offer guidance for designing new STWS materials that exploit this process. As shown in **Figure 7**, two defect properties are particularly critical: low vacancy formation energy and weak vacancy-vacancy interactions. The former enables high vacancy concentrations at temperatures relevant to STWS, while the latter allows for these vacancy concentrations to change substantially with respect to oxygen partial pressure. Together, these properties enable a large swing in vacancy concentration across the redox cycle and, consequently, a high $H_2$ yield.

Our findings also highlight the importance of antisite disorder in achieving low vacancy formation energies. This points to a more general materials design strategy for STWS: by placing redox-active cations on thermodynamically unfavorable (but kinetically accessible) sites, the cost of removing those cations (and thus forming vacancies) can be reduced. For this strategy to succeed, the cations must not only occupy high-energy configurations under oxidizing conditions but must also remain mobile enough to exchange with the surface during cycling. These considerations offer a conceptual framework for engineering cation-vacancy-based STWS materials beyond the iron aluminate spinel system studied here.

**Acknowledgements**

This work was supported by new faculty start-up funds from the University of Minnesota. The authors also acknowledge the Minnesota Supercomputing Institute (MSI) at the University of Minnesota for providing resources that contributed to the research results reported herein.

## Supplementary Information

## Cation vacancies mediate thermochemical water splitting with iron aluminates

Nathan J. Szymanski[1], Kent J. Warren[2], Alan W. Weimer[3], and Christopher J. Bartel[1,*]

[1] University of Minnesota, Department of Chemical Engineering and Materials Science, Minneapolis, MN 55455

[2] Solar Energy Engineering Laboratory, Department of Mechanical and Process Engineering, ETH Zürich, 8092 Zürich, Switzerland

[3] Department of Chemical and Biological Engineering, University of Colorado, Boulder, CO 80309

[*] Correspondence to cbartel@umn.edu

**Supplementary Note 1. Generation and quality of special quasirandom structures**

Special quasirandom structures (SQSs) were generated using the *sqsgenerator* package from Gehringer *et al.* (Comput. Phys. Commun. **286**, 108664, 2023), which employs a Monte Carlo optimization algorithm to approximate the short-range order of an ideal random solution within a finite supercell. For each inversion fraction in $FeAl_2O_4$ and $Fe_{1.5}Al_{1.5}O_4$, a 112-atom supercell was constructed containing 16 tetrahedral and 32 octahedral cation sites. During optimization, the nearest-neighbor pair correlations within each cation sublattice were iteratively adjusted to reproduce those of an ideal random distribution.

A total of $10^8$ Monte Carlo iterations were performed for each configuration. The degree to which each resulting SQS ($\sigma$) reproduces ideal randomness was quantified by the objective function:

$$O(\sigma) = \sum_{\xi,\eta} \left| \tilde{\alpha}_{\xi,\eta} - \alpha_{\xi,\eta}(\sigma) \right|$$

In this equation, $\tilde{\alpha}^{i}_{\xi,\eta}$ are the desired Warren-Cowley parameters of the random solution for pairs of species $\xi$ and $\eta$, whereas $\alpha_{\xi,\eta}(\sigma)$ are those of the generated SQS. Lower values are considered better, with $O(\sigma) = 0$ representing a perfect match with the random solution.

The resulting objective-function values for all iron aluminate SQSs are reported in **Table S1**. These range from 0 to 0.08, indicating reasonably close agreement with ideal randomness while maintaining computationally tractable supercell sizes. Octahedral and tetrahedral sublattices were optimized independently to preserve the correct inversion fraction on each sublattice.

**Supplementary Table 1:** Objective-function values $O(\sigma)$ quantifying the deviation of each generated SQS from ideal random pair correlations $FeAl_2O_4$ and $Fe_{1.5}Al_{1.5}O_4$ at different inversion fractions. Lower values indicate that configurations more closely reproduce the target Warren-Cowley parameters of an ideal random solution.

| | **$FeAl_2O_4$ Objectives** | | **$Fe_{1.5}Al_{1.5}O_4$ Objectives** | |
|---|---|---|---|---|
| **Inversion** | **Octahedral** | **Tetrahedral** | **Octahedral** | **Tetrahedral** |
| 6.25% | 0.0800 | 0.0714 | 0.0000 | 0.0714 |
| 12.50% | 0.0000 | 0.0256 | 0.0078 | 0.0256 |
| 18.75% | 0.0307 | 0.0000 | 0.0065 | 0.0000 |
| 25.00% | 0.0000 | 0.0182 | 0.0069 | 0.0182 |
| 31.25% | 0.0000 | 0.0667 | 0.0096 | 0.0667 |
| 37.50% | 0.0218 | 0.0159 | 0.0114 | 0.0159 |
| 43.75% | 0.0000 | 0.0000 | 0.0012 | 0.0000 |
| 50.00% | 0.0190 | 0.0000 | 0.0184 | 0.0000 |
| 56.25% | 0.0000 | 0.0159 | 0.0645 | 0.0159 |
| 62.50% | 0.0078 | 0.0667 | 0.0308 | 0.0667 |

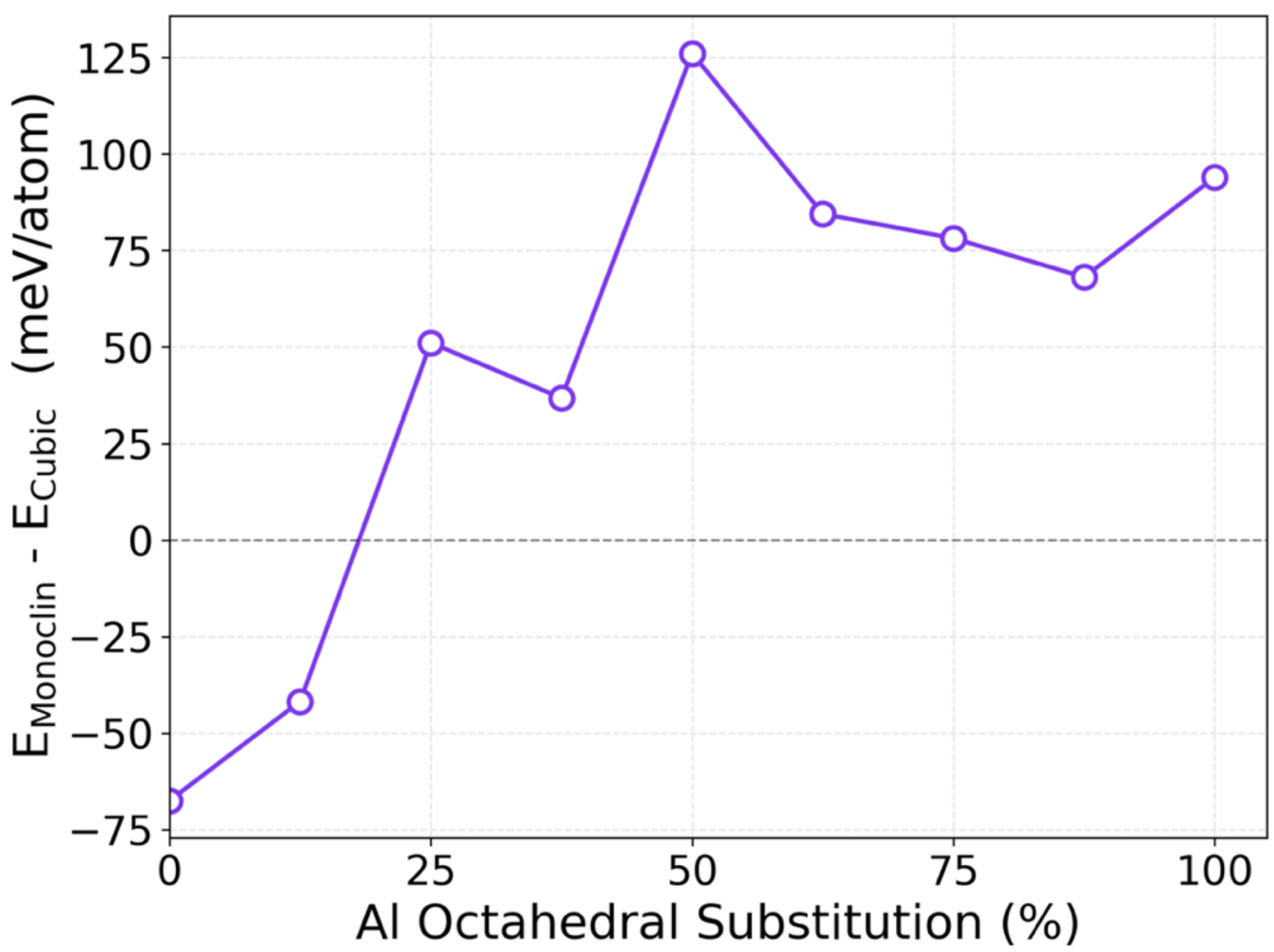

**Supplementary Figure 1:** Energy differences between cubic and monoclinic $Fe(Fe_{1-x}Al_x)_2O_4$ as a function of Al substituted on the octahedral sublattice. Energies were obtained from DFT calculations performed on special quasirandom structures, assuming Fe/Al disorder on the octahedral sites. All structures were relaxed to minimize energy with respect to cell volume while maintaining symmetry.

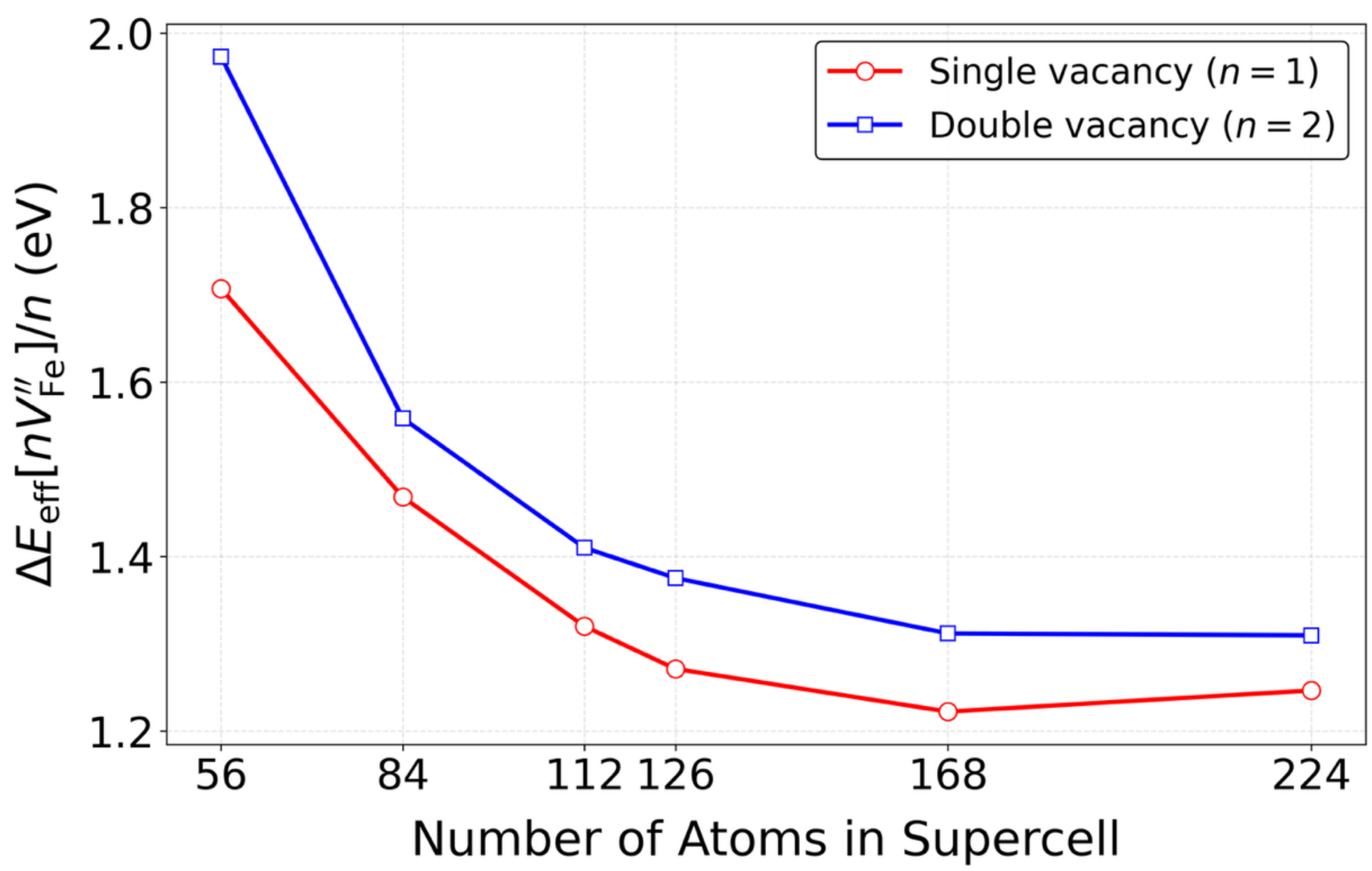

**Supplementary Figure 2:** Sensitivity of octahedral Fe vacancy formation energy in $Fe_3O_4$ to the size of the simulation cell. Shown is the effective per-vacancy formation energy, $\Delta E_{\mathrm{eff}}\,[\mathrm{V}''_{\mathrm{Fe}}]/n$, computed at 1400 °C for increasingly larger cells. Relatively small changes in vacancy formation energies are observed at cell sizes > 112 atoms, and we therefore used this cell size throughout our remaining calculations.

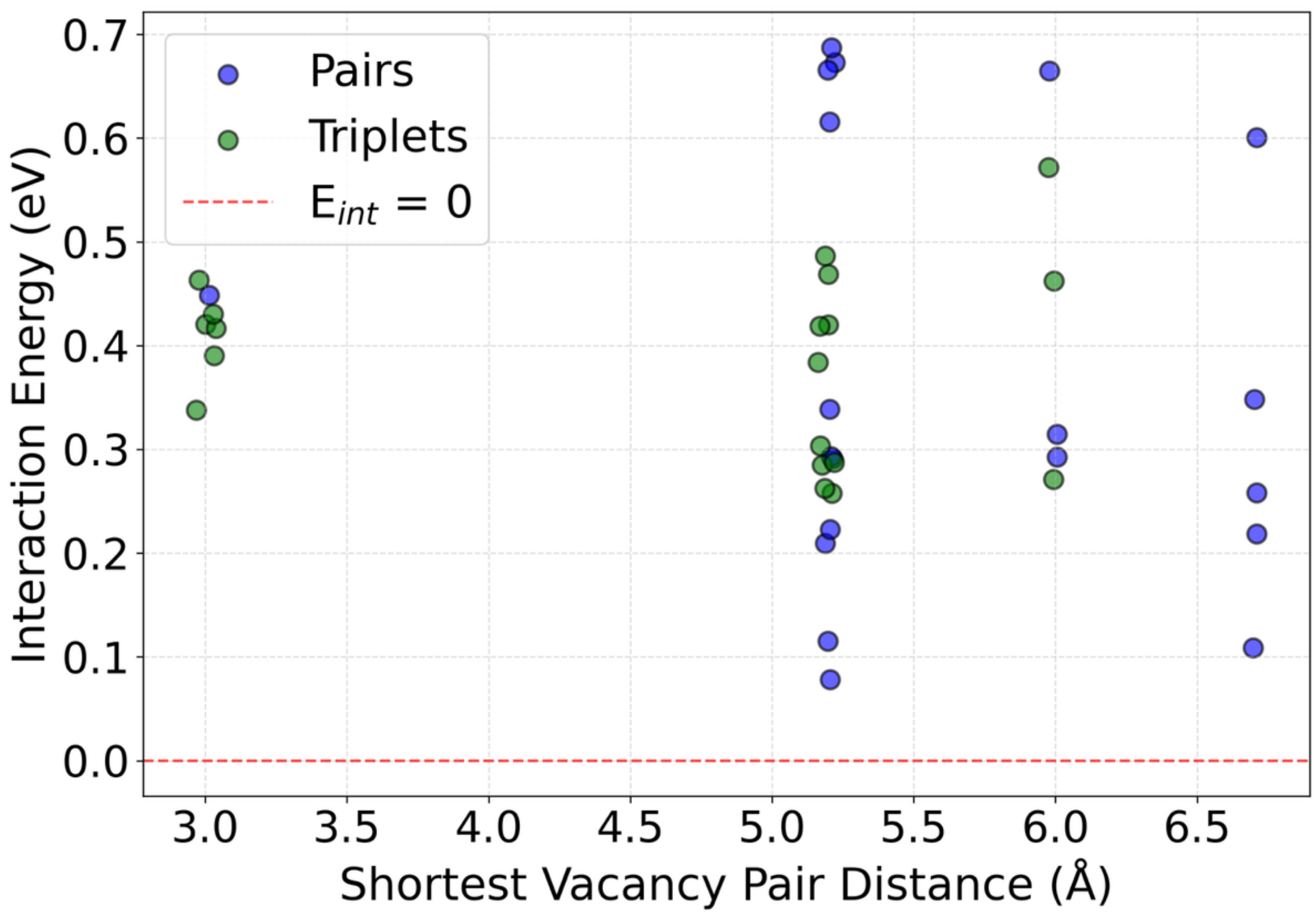

**Supplementary Figure 3:** Vacancy-vacancy interaction energy ($E_{\mathrm{int}}$) versus shortest Fe-Fe separation for di- and tri-vacancy configurations in $Fe_3O_4$, computed relative to the minimum single-vacancy 0 K formation energy at 1400 °C. Each point corresponds to one randomly sampled configuration, and the solid horizontal line denotes $E_{\mathrm{int}} = 0$. The blue dots (pairs) are energies from supercells containing two vacancies, while the green dots (triplets) are from supercells with three vacancies.

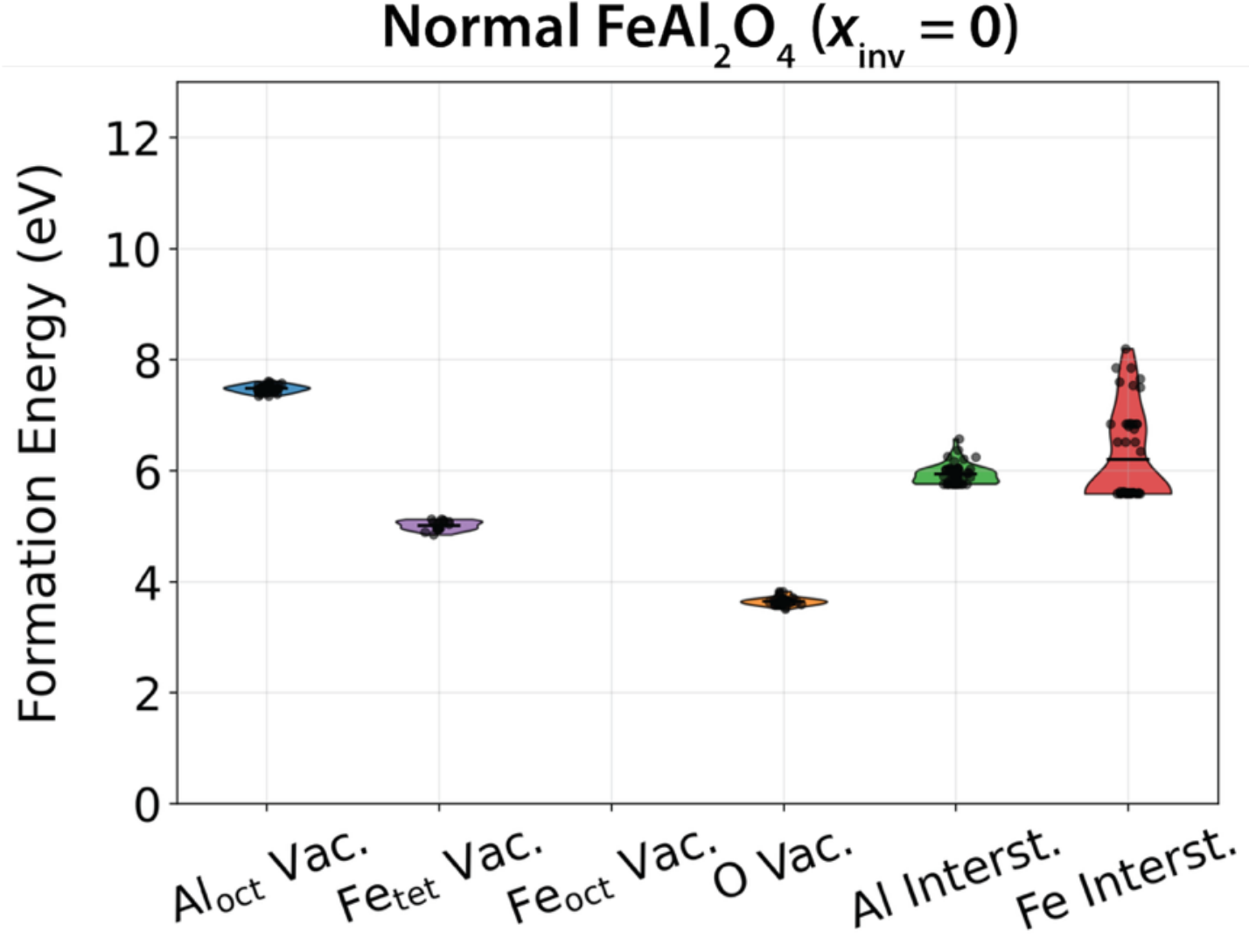

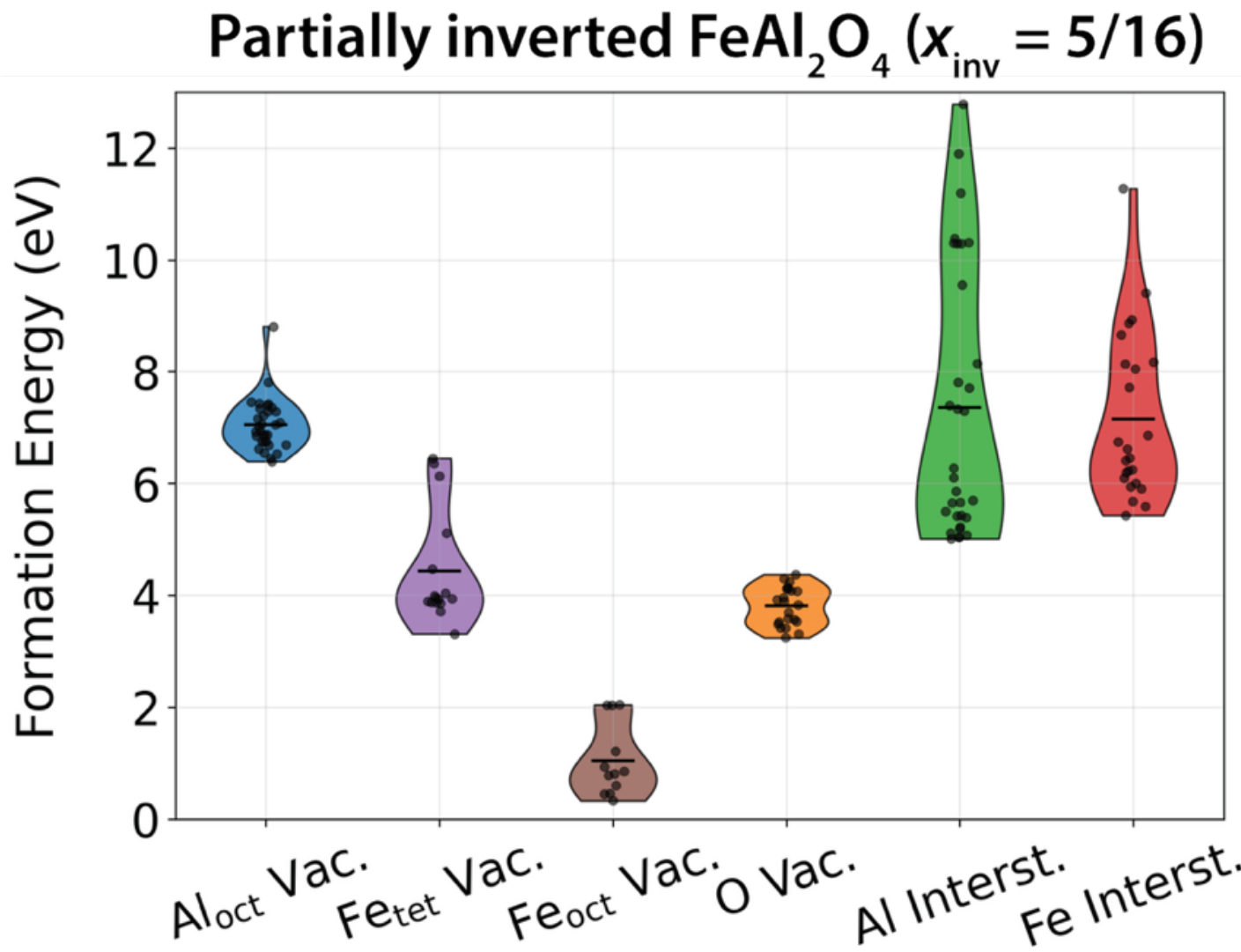

**Supplementary Figure 4:** Formation energies of isolated point defects in $FeAl_2O_4$ with (top) and without (bottom) cation inversion. These were computed on 112-atom supercells. For the vacancy data, each point represents the formation energy of a single configuration with one random site removed. Octahedral Fe vacancies are not shown in the top subplot since the normal spinel (without inversion) has 0% Fe occupancy on the octahedral sublattice. For the interstitial data, viable sites were identified by scanning Voronoi voids and keeping positions at least one ionic radius of the inserted species away from closest atoms. All calculations were performed keeping the unit cell neutral and its shape fixed, while allowing its volume to relax. The chemical potentials used in computing the defect formation energies are: $\mu_{\mathrm{Fe}} = -6.55$ eV, $\mu_{\mathrm{Al}} = -3.75$ eV, and $\mu_{\mathrm{O}} = -7.47$ eV. The Fe and Al chemical potentials were obtained from DFT energies of bulk Fe (BCC) and Al (FCC) in their ground states, with a +U correction applied to Fe (Wang *et al.*, Sci. Rep. **11**, 15496, 2021). The oxygen chemical potential was determined from DFT calculations on an $O_2$ dimer, including a gas-phase correction corresponding to $T = 1400$ °C and $p_{\mathrm{O_2}} = 10^{-5}$ bar.

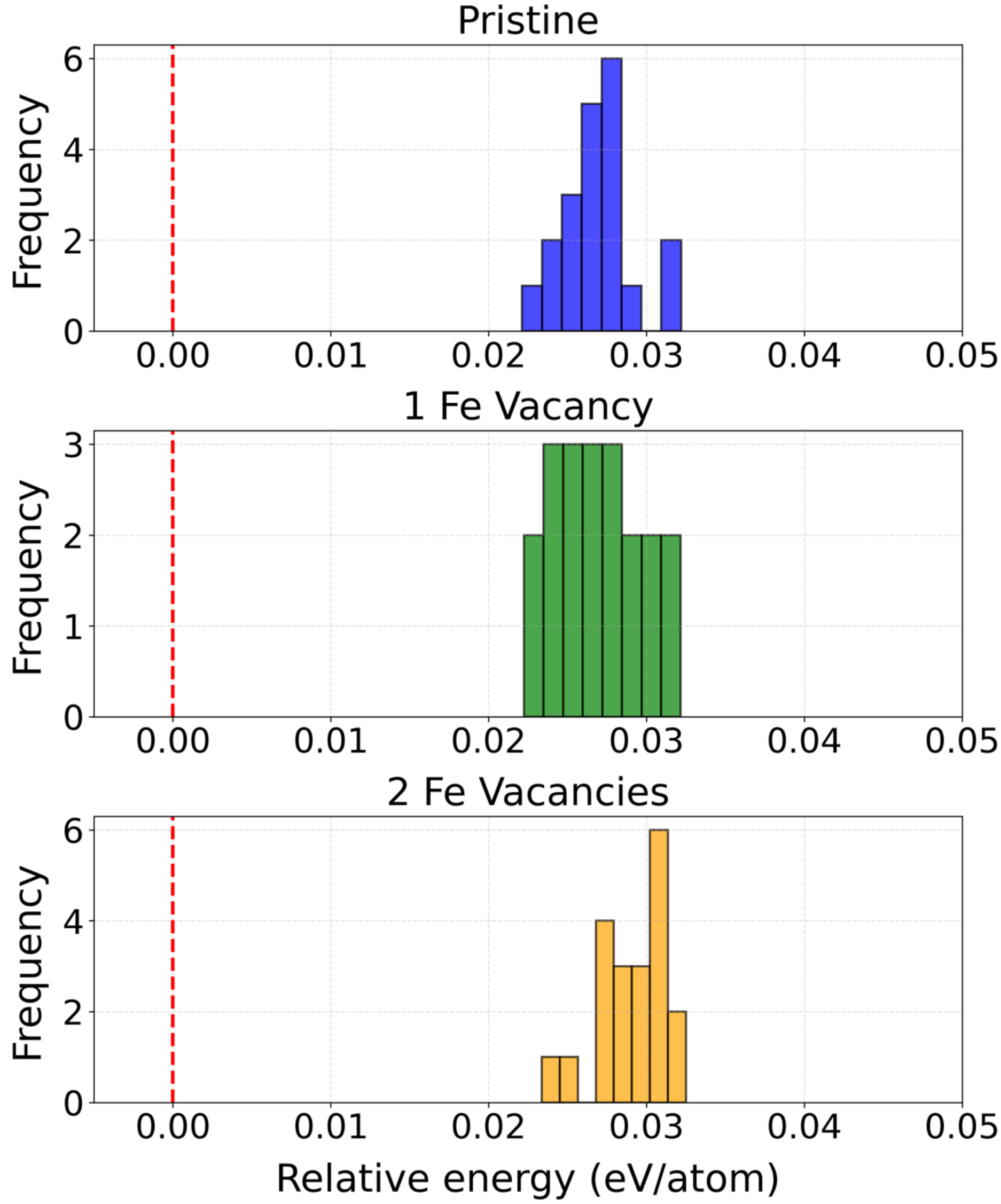

**Supplementary Figure 5:** Distribution of energies for $Fe_3O_4$ structures with different random spin initializations for pristine (top), single-vacancy (middle), and double-vacancy (bottom) configurations. These are plotted relative to our computed ground state (red dashed line, set to zero).

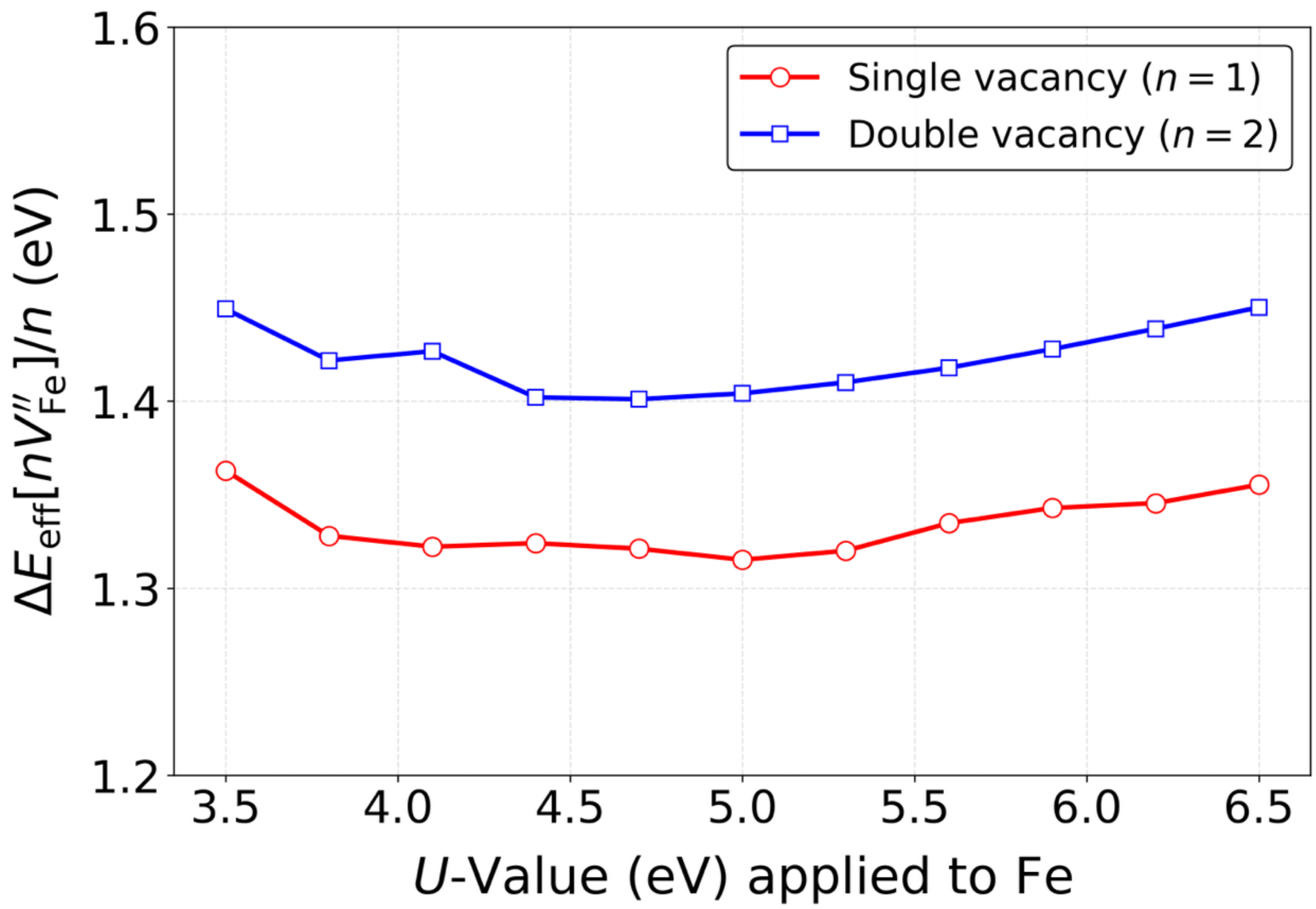

**Supplementary Figure 6:** Sensitivity of octahedral Fe vacancy formation energy in $Fe_3O_4$ to the Hubbard $U$ applied to Fe 3d states. Shown is the effective per-vacancy formation energy, $\Delta E_{\mathrm{eff}}\,[\mathrm{V}''_{\mathrm{Fe}}]/n$, computed at 1400 °C for supercells with one ($n = 1$) and two ($n = 1$) vacancies as a function of $U$. The $U$ -dependence is weak (varying < 0.1 eV) across the range of values tested (3.5 to 6.5 eV); we chose to use $U$ = 5.3 eV in all remaining calculations.